\documentclass[]{aastex631}

\usepackage{xspace}
\usepackage{natbib}
\usepackage{multirow}

\newcommand{\threeml}{\texttt{3ML}}

\def\arcsec{\hbox{$^{\prime\prime}$}}

\def\lae{\mathrel{\raise .4ex\hbox{\rlap{$<$}\lower 1.2ex\hbox{$\sim$}}}}
\def\gae{\mathrel{\raise .4ex\hbox{\rlap{$>$}\lower 1.2ex\hbox{$\sim$}}}}

\newcommand{\ixpe}{{IXPE}\xspace}

\newcommand{\Rm}[1]{\uppercase\expandafter{\romannumeral #1\relax}}

\received{October 5, 2021}
\received{\today}
\submitjournal{Astrophysical Journal}

\shorttitle{Observations of LSP \& ISP Blazars with IXPE}

\shortauthors{Marshall et al.}
\begin{document}

\title{Observations of Low and Intermediate Spectral Peak Blazars with the Imaging X-ray Polarimetry Explorer}

%% you desire. In v6.31 you can include a footnote in the title.

\correspondingauthor{Herman L.\ Marshall}
\email{hermanm@mit.edu}

\author[0000-0002-6492-1293]{Herman L. Marshall}
\affiliation{MIT Kavli Institute for Astrophysics and Space Research, Massachusetts Institute of Technology, 77 Massachusetts Avenue, Cambridge, MA 02139, USA}

%%%% directly involved Tier 1 authors

\author[0000-0001-9200-4006]{Ioannis Liodakis}
\affiliation{Finnish Centre for Astronomy with ESO, 20014 University of Turku, Finland}

\author[0000-0001-7396-3332]{Alan P. Marscher}
\affiliation{Institute for Astrophysical Research, Boston University, 725 Commonwealth Avenue, Boston, MA 02215, USA}
\author[0000-0002-7574-1298]{Niccol\`o Di Lalla}
\affiliation{Department of Physics and Kavli Institute for Particle Astrophysics and Cosmology, Stanford University, Stanford, California 94305, USA}

\author[0000-0001-9522-5453]{Svetlana G. Jorstad}
\affiliation{Institute for Astrophysical Research, Boston University, 725 Commonwealth Avenue, Boston, MA 02215, USA}
\affiliation{St. Petersburg State University, St. Petersburg, 199034 Russia}

\author[0000-0001-5717-3736]{Dawoon E.\ Kim}
\affiliation{INAF Istituto di Astrofisica e Planetologia Spaziali, Via del Fosso del Cavaliere 100, 00133 Roma, Italy}
\affiliation{Dipartimento di Fisica, Universit\`{a} degli Studi di Roma ``La Sapienza'', Piazzale Aldo Moro 5, 00185 Roma, Italy}
\affiliation{Dipartimento di Fisica, Universit\`{a} degli Studi di Roma ``Tor Vergata'', Via della Ricerca Scientifica 1, 00133 Roma, Italy}

\author[0000-0001-9815-9092]{Riccardo Middei}
\affiliation{Space Science Data Center, Agenzia Spaziale Italiana, Via del Politecnico snc, 00133 Roma, Italy}
\affiliation{INAF Osservatorio Astronomico di Roma, Via Frascati 33, 00040 Monte Porzio Catone (RM), Italy}

\author[0000-0002-6548-5622]{Michela Negro}
\affiliation{University of Maryland, Baltimore County, Baltimore, MD 21250, USA}
\affiliation{NASA Goddard Space Flight Center, Greenbelt, MD 20771, USA}
\affiliation{Louisiana State University, Baton Rouge, LA 70803, USA}

\author[0000-0002-5448-7577]{Nicola Omodei}
\affiliation{Department of Physics and Kavli Institute for Particle Astrophysics and Cosmology, Stanford University, Stanford, California 94305, USA}

\author[0000-0001-6292-1911]{Abel L. Peirson}
\affiliation{Department of Physics and Kavli Institute for Particle Astrophysics and Cosmology, Stanford University, Stanford, California 94305, USA}

% ...



\begin{abstract}
We present X-ray polarimetry observations from the Imaging X-ray Polarimetry Explorer (IXPE) of three low spectral peak
and one intermediate spectral peak
blazars, namely 3C~273, 3C~279, 3C~454.3, and S5~0716$+$714. For none of these objects was \ixpe\ able to detect X-ray polarization at the 3$\sigma$ level. However, we placed upper limits on the polarization degree at $\sim$10-30\%. The undetected polarizations favor models where the X-ray band is dominated by unpolarized photons upscattered by relativistic electrons in the jets of blazars, although hadronic models are not completely eliminated. We discuss the X-ray polarization upper limits in the context of our contemporaneous multiwavelength polarization campaigns. 

\end{abstract}

%\keywords{}

\section{Introduction}
\label{sec:intro}

Blazars are active galactic nuclei (AGN) with relativistic jets that are inferred to
be oriented within a few degrees from the observer's line of sight \citep[cf.][]{2019ARA&A..57..467B}.
Their spectral energy distributions (SEDs) are generally characterized by two emission components: one at low photon energies (radio through optical, UV, and, in some cases, X-ray bands) and the other at much higher (X-ray and $\gamma$-ray) energies.
Blazars are often classified by the frequency at which the low-energy portion of the SED peaks.  
This component is modeled as electron synchrotron emission. 
When the peak frequency is below the optical or infrared (IR) band, the blazar is considered to be low-synchrotron-peaked (LSP), in contrast with
high-synchrotron-peaked (HSP) blazars, whose peak is in the X-ray band. Objects with a peak in the optical/UV range are termed intermediate-synchrotron-peaked (ISP) blazars.

Several LSP/ISP blazars were observed in 2022 with the Imaging X-ray Polarimetry Explorer \citep[{\ixpe,}][]{Weisskopf2022} that was launched on 2021 December 9. HSP blazars such as Mrk 501 \citep{2022Natur.611..677L}, Mrk 421 \citep{2022ApJ...938L...7D}, and PG 1553$+$113 \citep{Middei_2023} were measured by \ixpe\ to be polarized in the 10-15\% range, 2--5 times higher than the optical polarization at the same epochs. These findings support a model by which the X-rays in HSPs are produced in energy-stratified shocks, with
electrons accelerated at the shock front to energies high enough to emit X-ray synchrotron radiation. Such energetic electrons occur only in a thin layer close to the shock front, beyond which the maximum electron energy decreases owing to radiative cooling \citep{MarscherGear1985}. Consequently, the optically emitting region, which requires somewhat lower electron energies, extends over a more extensive volume.
If the magnetic field is mostly disordered, e.g., by turbulence, then the mean polarization is expected to be stronger and more highly variable over the smaller volume from which the X-rays arise \citep{DiGesu2022}.
The ratio of X-ray to optical polarization was observed to be higher than expected in such a model, which suggests that the level of disorder of the magnetic field increases with distance from the shock front \citep{2022Natur.611..677L,MarscherJorstad2022}.
The alignment of the electric vector position angle (EVPA) with the jet direction in Mrk 501 \citep{2022Natur.611..677L} implies that there is an ordered component to the magnetic field perpendicular to the jet. How this component arises is somewhat uncertain; various suggestions involve compression of the disordered field at the shock \citep{Hughes1985}, generation of a magnetic field component transverse to the shock normal \citep{2018MNRAS.480.2872T}, or a toroidal magnetic field component in the ambient jet \citep{2005MNRAS.360..869L}.  Complicating the physical picture of HSP blazar shocks are the observations that the EVPA can rotate in the X-ray band without rotation in the optical \citep{DiGesu2023} or vice versa \citep{Middei_2023}.

On the other hand, the X-ray portions of the SEDs of LSP and ISP blazars generally have flatter spectra than at optical frequencies \citep[e.g.,][]{Abdo2010}, indicating that a single synchrotron model is not responsible for both bands.
The flatter X-ray spectra are consistent with models where the X-ray band is dominated by Compton upscattering of low frequency (e.g. infrared) photons by the relativistic electrons generating the synchrotron radiation \citep[e.g.,][]{Maraschi2008}.
Processes involving hadrons, such as proton synchrotron radiation or production of electron-positron pairs by particle cascades might also be involved in generating the X-rays \citep[e.g.,][]{Zhang2016}. Each of these processes has specific polarization properties that can potentially identify their level of importance in the X-ray emission \citep[e.g.,][]{Liodakis2019,Peirson2022}.
Compton scattering of unpolarized seed photons, such as those produced by the broad emission-line region, produces essentially unpolarized high-energy photons \citep{Krawczynski2012}.
If the seed photons are from synchrotron radiation in the jet --- the synchrotron self-Compton (SSC) case --- the X-ray polarization fraction ($\Pi_X$) is
generally less than half
that of the synchrotron radiation; the exact ratio depends on the model and the level of disorder of the magnetic field \citep{Peirson2022}. For hadronic processes, $\Pi_X$ is expected to be similar or even exceed the optical polarization ($\Pi_O$) \citep{Zhang2013,Zhang2019}. Hence, X-ray polarization measurements or even upper limits in the context of simultaneous multiwavelength polarization observations can test emission models for LSP and ISP blazars.

\ixpe\ previously observed BL Lac on three occasions. In the first two observations, the X-ray spectrum of BL Lac resembled that of LSPs, in the third observation that of ISPs. During the first two observations, there was only a 99\% confidence upper limit of $\Pi_X < 13$\%, while the contemporaneous $\Pi_O$ exceeded the X-ray limits \citep{Middei2023}. In the third observation, 22\% polarization was detected in the 2--4 keV band during a time dominated by synchrotron-emitting electrons from the low-energy hump of the SED \citep{Peirson2023}.  There are still only upper limits to the X-ray polarization from the high-energy component. Those results disfavor any significant contribution from protons to the emission.

In this study, we present results from \ixpe\ observations of three LSP blazars and one ISP blazar that were observed in 2022: 3C~273, 3C~279, 3C~454.3, and S5~0716$+$714. 
In section \ref{sec:ixpe} we describe the \ixpe\ observations with full-band polarimetric and spectropolarimetric analyses, in section \ref{sec:notes} we discuss the multiwavelength polarization campaigns on individual sources, and in section \ref{sec:disc} we discuss our results.

\section{IXPE Observations}
\label{sec:ixpe}

A log of \ixpe\ observations is provided in Table~\ref{tab:observations}.  See \citet{Weisskopf2022} for details about \ixpe. At the $\sim30''$ angular resolution of \ixpe, all of these sources are spatially unresolved.

\subsection{Full-Band Analyses}
\label{sec:ixpe_full}

\begin{figure}
\centering
 \includegraphics[width=8.cm]{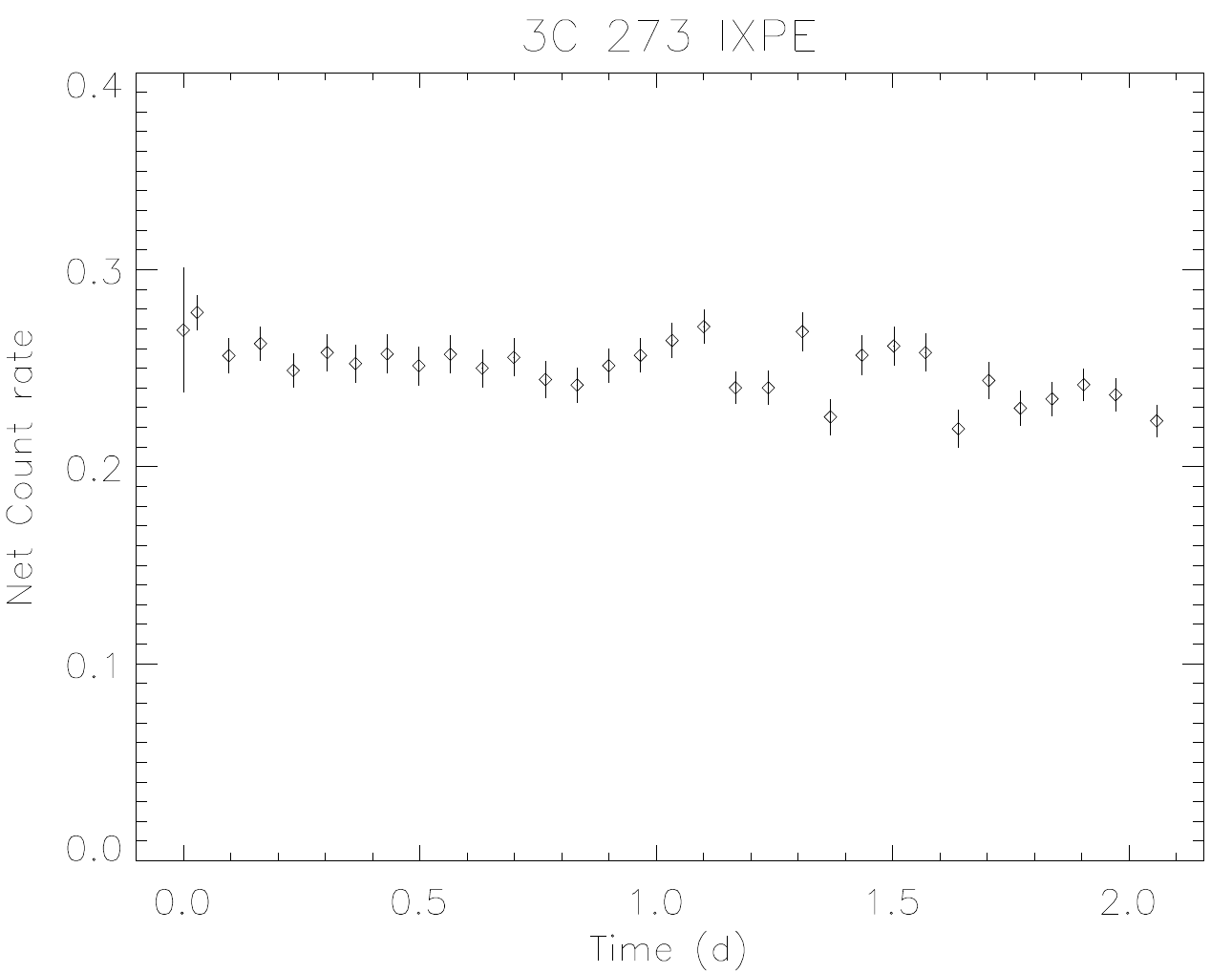}
 \includegraphics[width=8.cm]{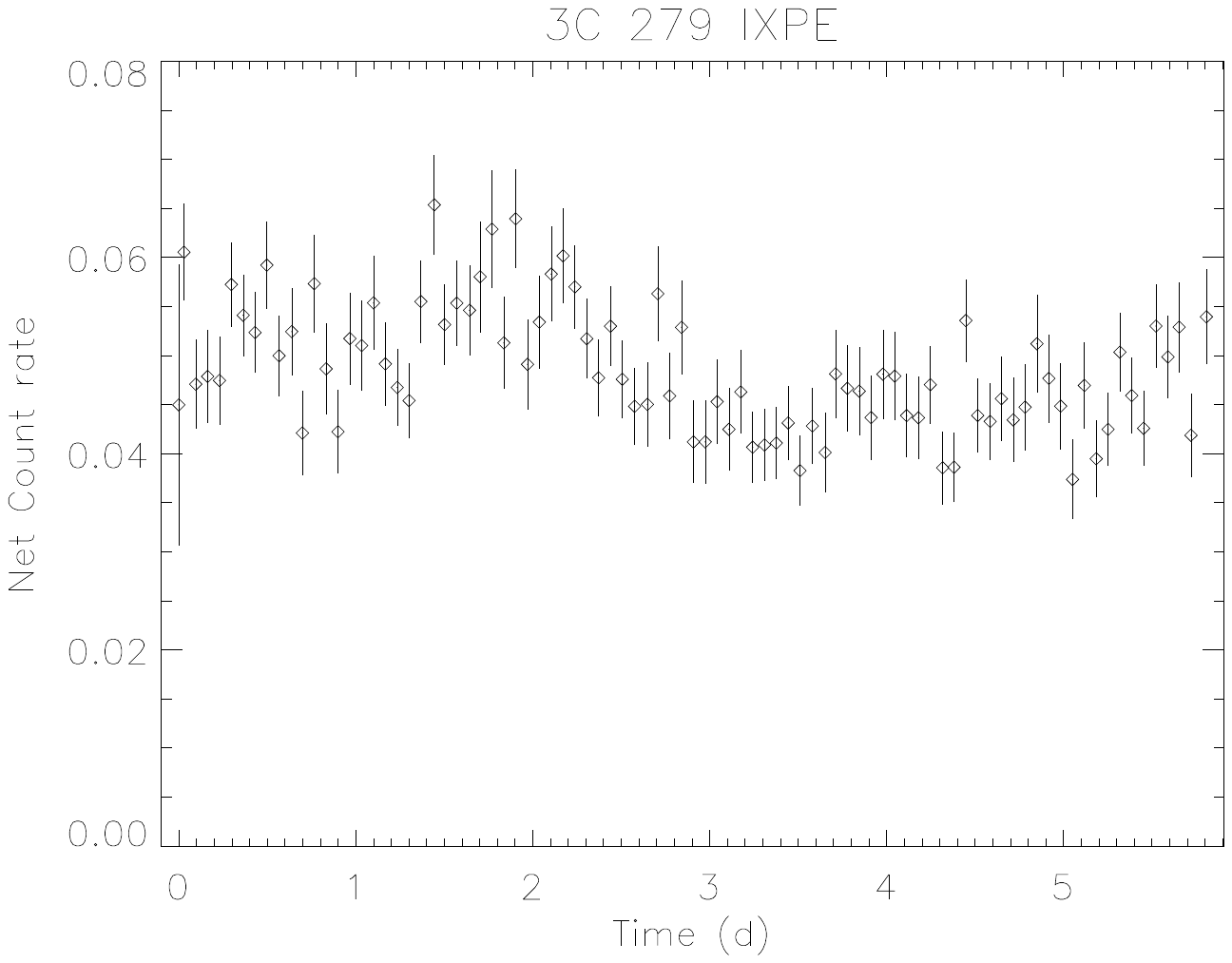}\\
 \includegraphics[width=8.cm]{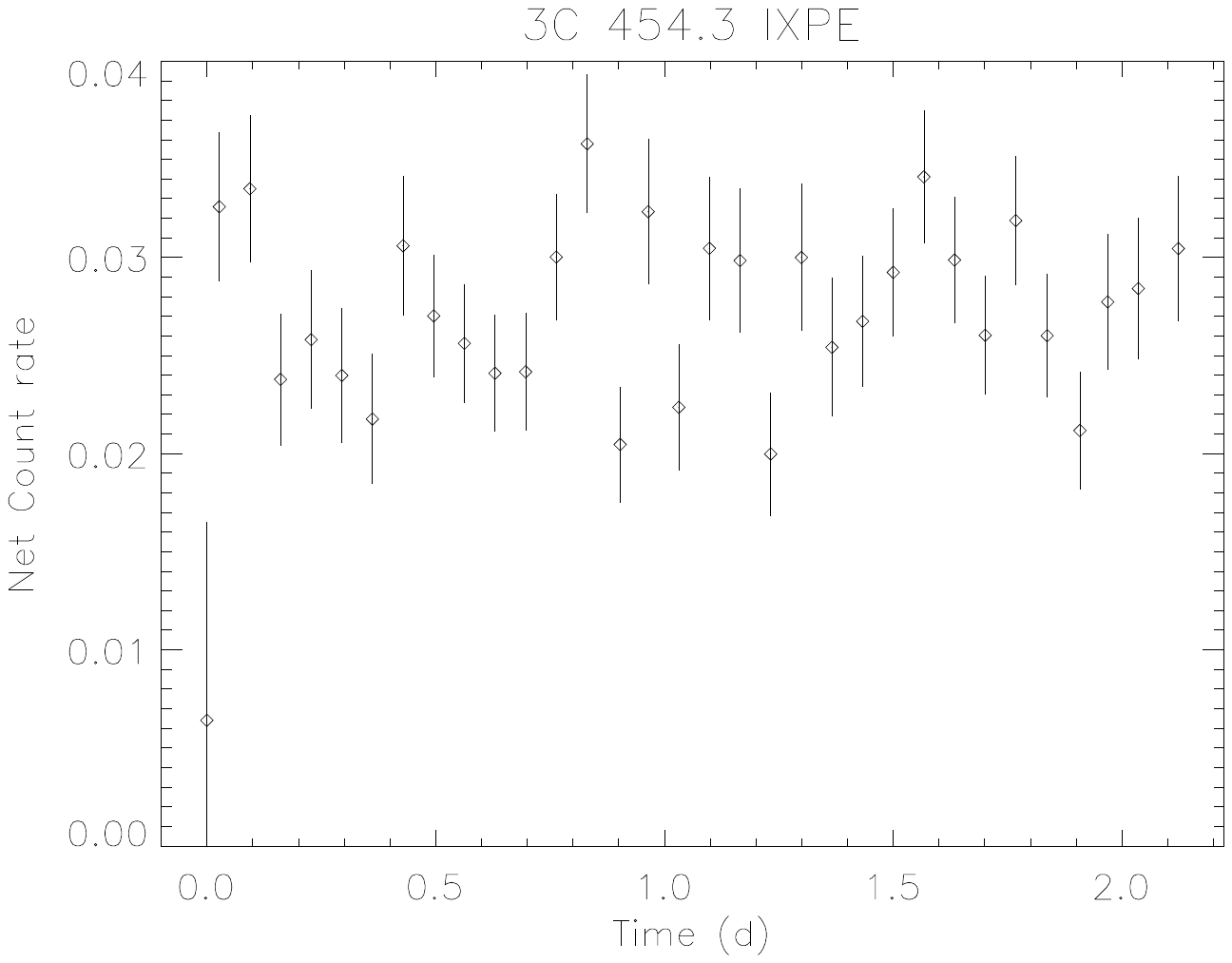}
 \includegraphics[width=8.cm]{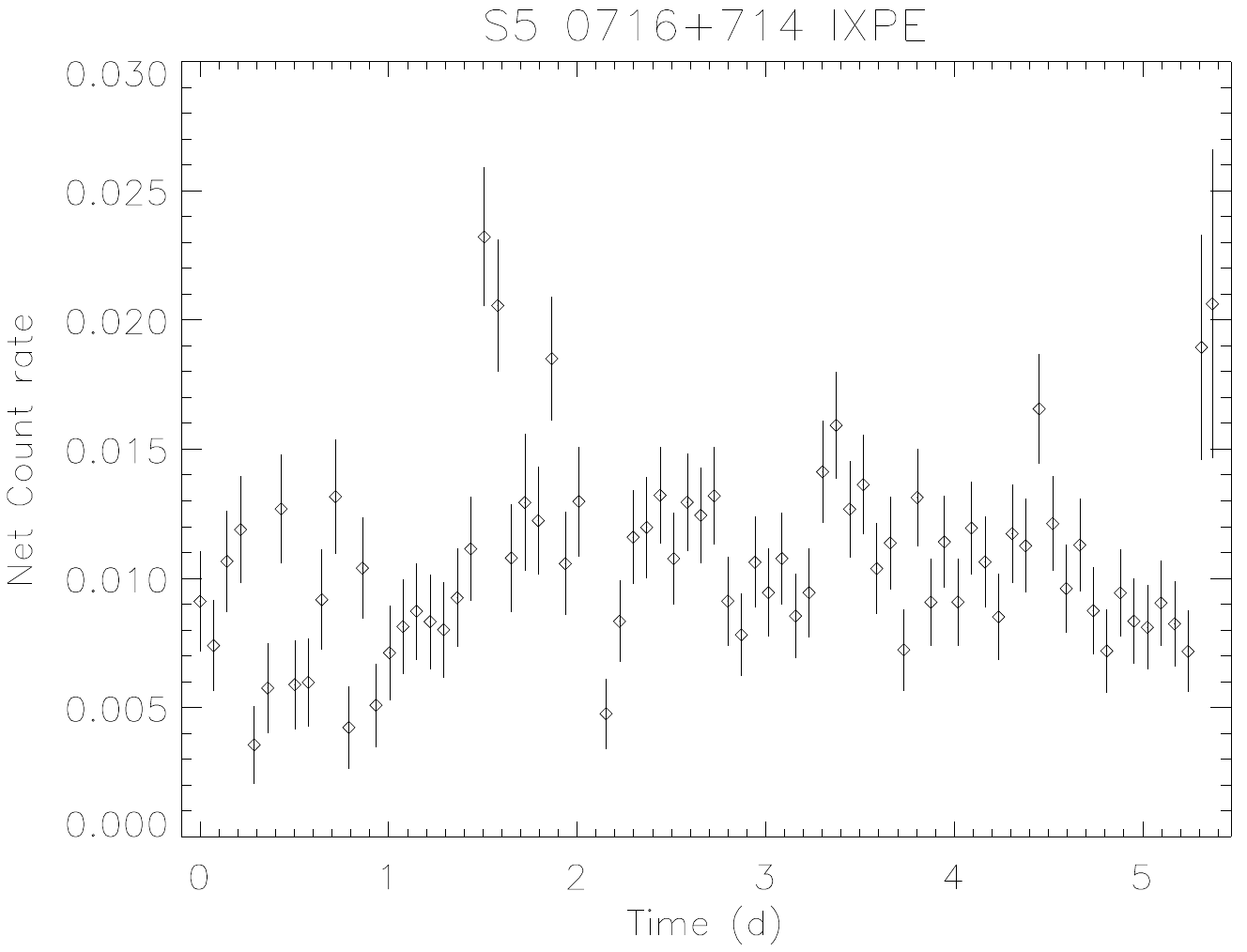}
 \caption{IXPE background-subtracted light curves for the four blazars reported here.  No large variations are observed that would warrant isolating specific time periods for detailed analysis.}
\label{fig:lightcurves}
\end{figure}

Event data were selected over the full bandpass from 2 to 8 keV, from a circular region 60\arcsec\ radius about each source, with background selected from an annulus from 200\arcsec\ to 300\arcsec\ from the target. Data were processed using {\tt ixpeobssim} \citep{2022SoftX..1901194B} and standard {\tt ftools} \cite{heasoft}.
Background-subtracted light curves are shown in Fig.~\ref{fig:lightcurves}.

\begin{figure}
\centering
 \includegraphics[width=8.cm]{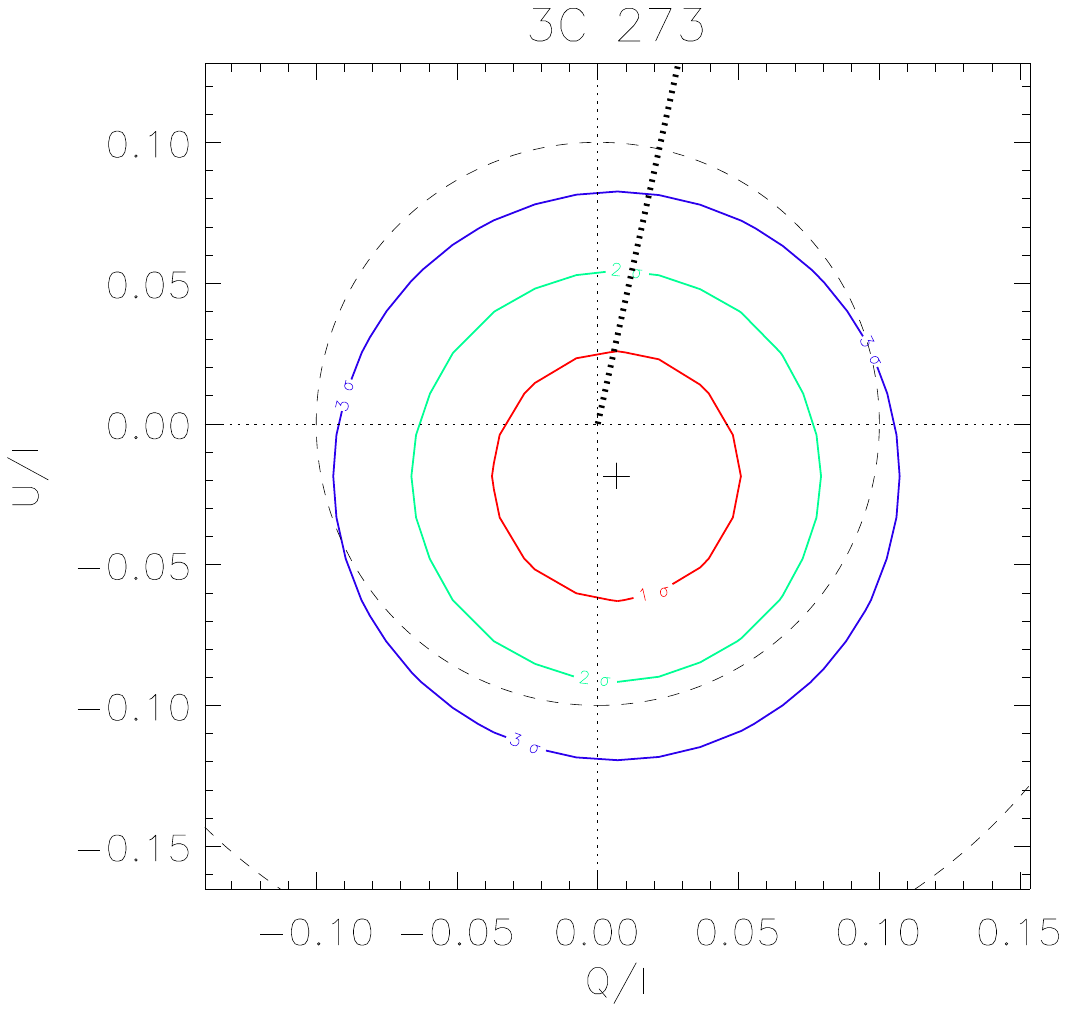}
 \includegraphics[width=8.cm]{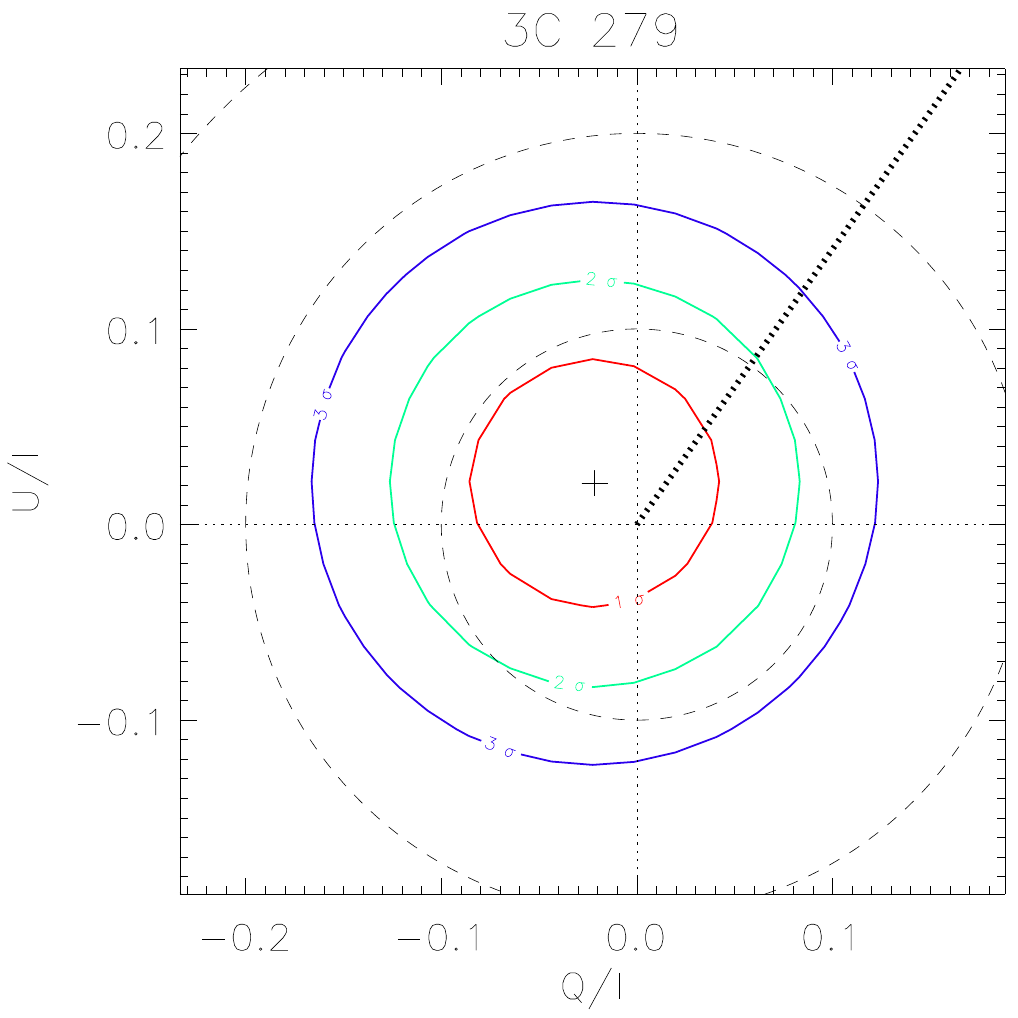}\\
 \includegraphics[width=8.cm]{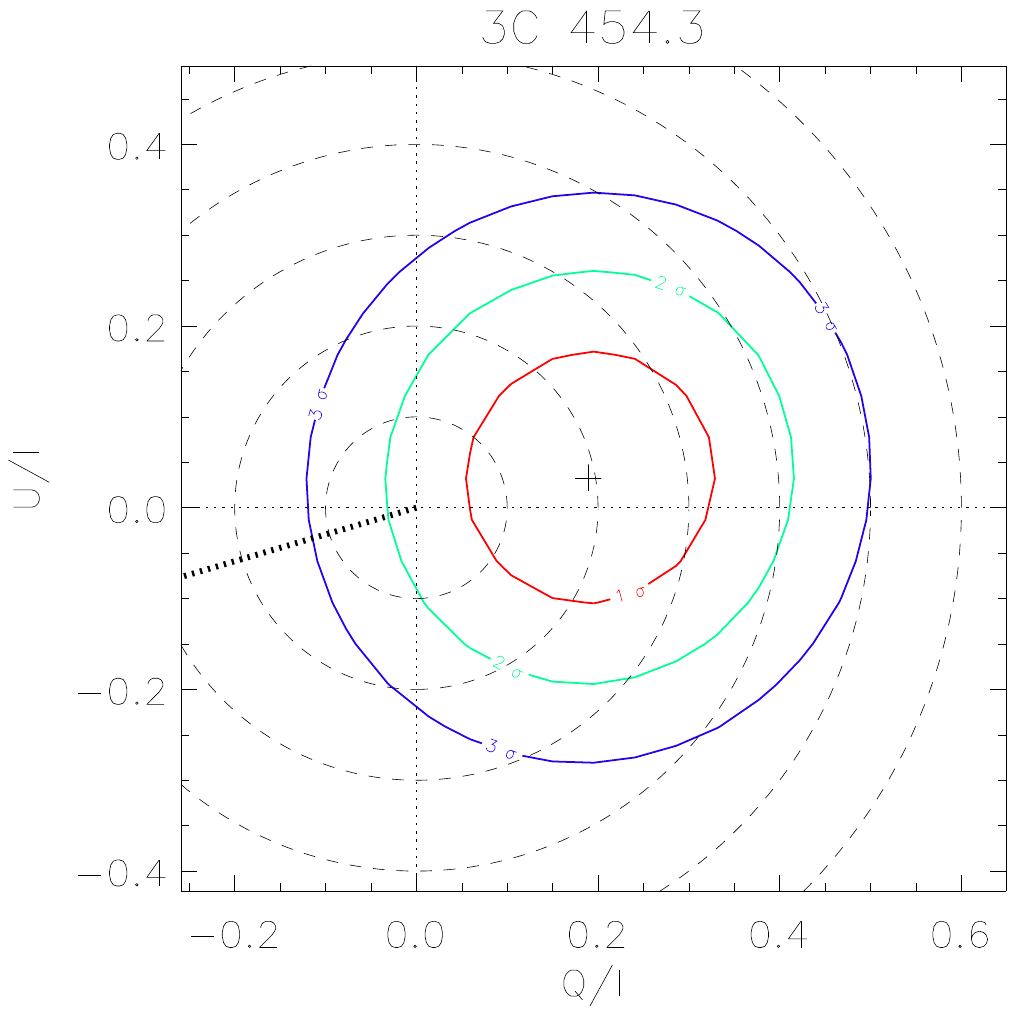}
 \includegraphics[width=8.cm]{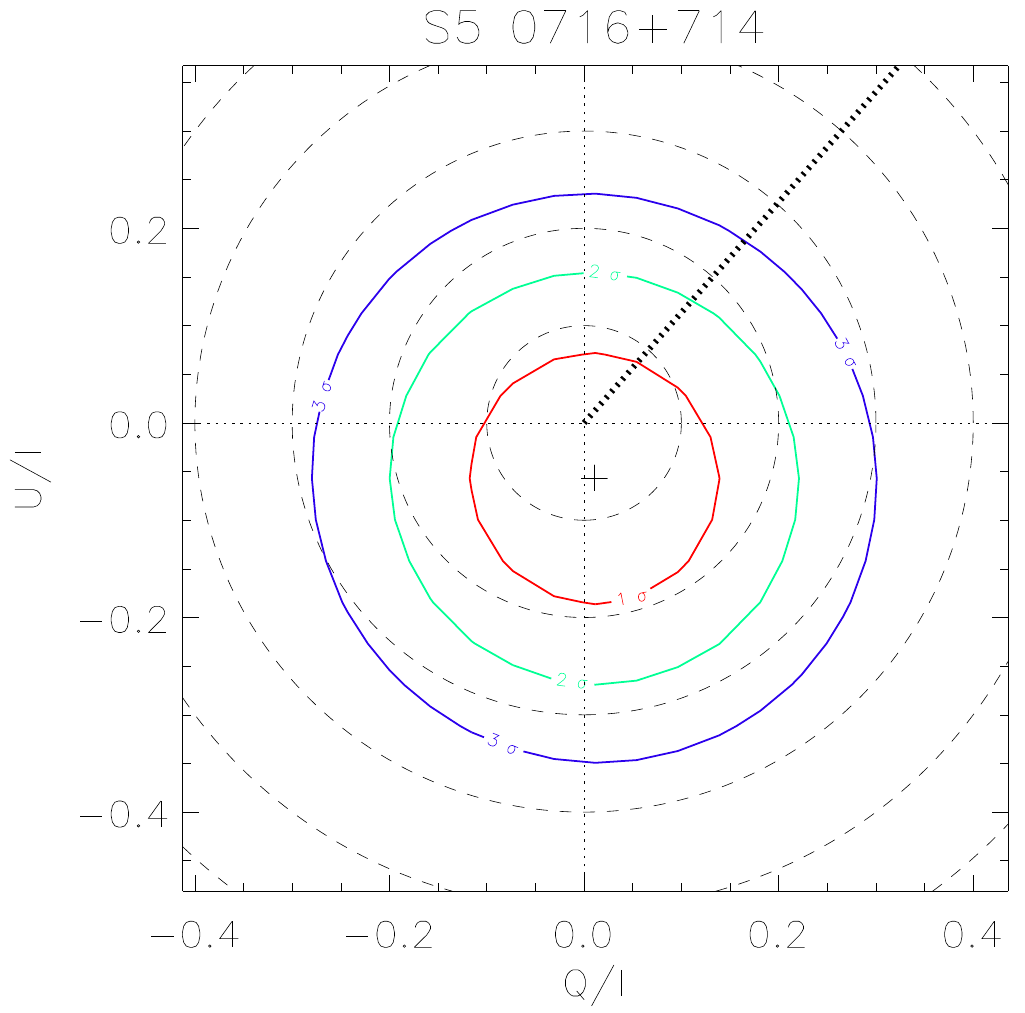}
 \caption{Probability contours in $Q-U$ space for time- and bandpass-averaged IXPE polarimetry
 of four blazars using a maximum likelihood method that accounts for unpolarized background.  Contour levels for two degrees of freedom enclose the 1$\sigma$, 2$\sigma$, and 3$\sigma$ confidence regions.  The plus signs mark the best estimates of $q = Q/I$ and $u = U/I$ and the dashed lines give circles of constant X-ray polarization fraction $\Pi_X = (q^2 + u^2)^{1/2}$ in 10\% increments.  The PAs of the jets are indicated by the thick dotted lines, taken from .
 In all cases, the results are consistent with null polarization at the 1--2$\sigma$ confidence level.}
\label{fig:qu_lsps}
\end{figure}

Results from an unbinned event-based likelihood analysis \citep{marshall20} that accounts for an unpolarized background \citep{Marshall2023} are shown in Fig.~\ref{fig:qu_lsps}.  No clear detections of X-ray polarizations are observed at the 1--2$\sigma$ confidence level, with $99$\% confidence limits ranging from 9\% to 28\% (Table~\ref{tab:observations}).

\begin{figure}
 \centering
 \includegraphics[width=8cm]{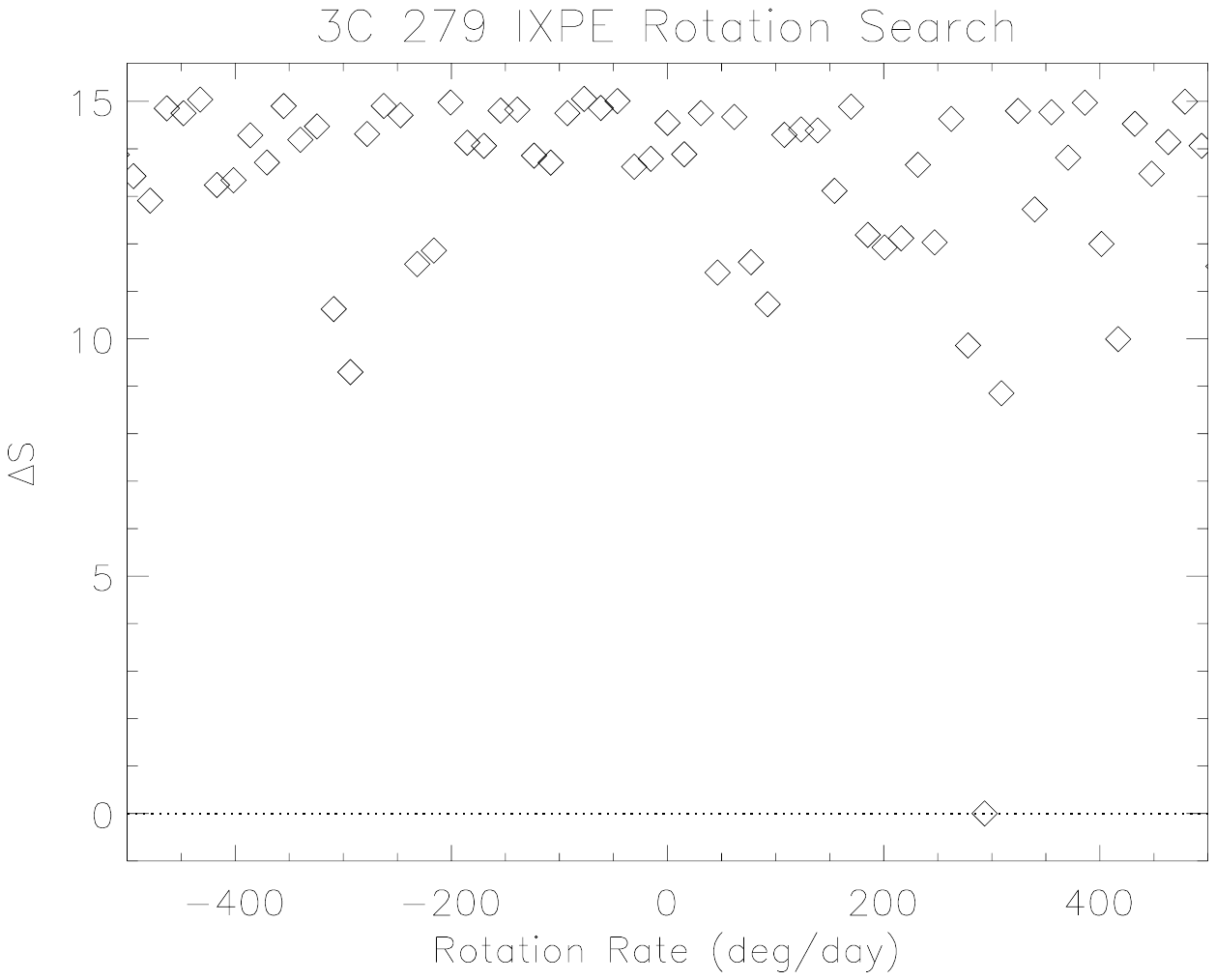}
 \caption{Example of a search for EVPA rotation, as applied to 3C~279.  This source provided the largest log-likelihood difference of this group of blazars.  A value of $\Delta S$ as large as 15.05 is significant at the 99.95\% confidence level but there were about 130 trial rates tested, giving a chance detection of a rotation of this significant of 6.8\%.  See \cite{DiGesu2023} for details of how the test was conducted.}
\label{fig:rotsearch}
\end{figure}

Rotations of the EVPA during the IXPE observation, especially for exposures longer than the typical variability timescales ($\sim$1--2 days) in blazars, can lead to depolarization, as was demonstrated recently in the X-ray EVPA rotation detected by IXPE in Mrk~421 \citep{DiGesu2023}.
We tested for rotations using the method described in \cite{DiGesu2023}, employing an event-based likelihood method \citep{marshall20}.  Briefly, event tracks are rotated according to a simple model where the EVPA rotates uniformly through the observation with rate $\omega$ (in degrees per day).  The difference in the log-likelihood is $\Delta S = S(\omega; \hat{q}, \hat{u}) - S(0; \hat{q_0}, \hat{u_0})$, where $S(\omega; \hat{q}, \hat{u})$ is the log likelihood at $\omega$ at the best fit $q$ and $u$ for that value of $\omega$.  $\Delta S$ is distributed as $\chi^2$ with one degree of freedom because $q$ and $u$ are uninteresting parameters for this test.
We do not find evidence for EVPA rotations during our observations in any of the four sources.
An example of the results from a rotation search is shown in Fig.~\ref{fig:rotsearch}.

\begin{deluxetable}{rrrcrc}
\caption{Summary of IXPE Observations}
\tablehead{
\colhead{Source} & \colhead{Instrument} & \colhead{Observation ID} & \colhead{MJD range} & \colhead{Exposure (ks)\tablenotemark{a}} & \colhead{$\rm \Pi_X$\tablenotemark{b}}\\[-0.2 cm]
}
\startdata
3C 273 & \ixpe & 01005901 & 59732.37 - 59734.45 &  95.28 & $< 9.0$\%  \\
3C 279 & \ixpe & 01005701 & 59743.02 - 59748.85 & 264.42 & $< 12.7$\%\\
3C 454.3 & \ixpe & 01005401 & 59730.19 - 59732.34 &  98.12 & $< 28$\%\\
S5 0716+714 & \ixpe & 01005301 & 59669.43 - 59674.80 & 358.68 &  $< 26$\%\\
\enddata
\tablenotetext{a}{Average of exposures for the three detector units.}
\tablenotetext{b}{99\% confidence limits using the unbinned, event-based likelihood method (\S~\ref{sec:ixpe_full}).}
\label{tab:observations}   
\end{deluxetable}

\subsection{Spectropolarimetry}
\label{sec:ixpe_spectr}

\begin{figure}[b]
\centering
 \includegraphics[width=8.cm]{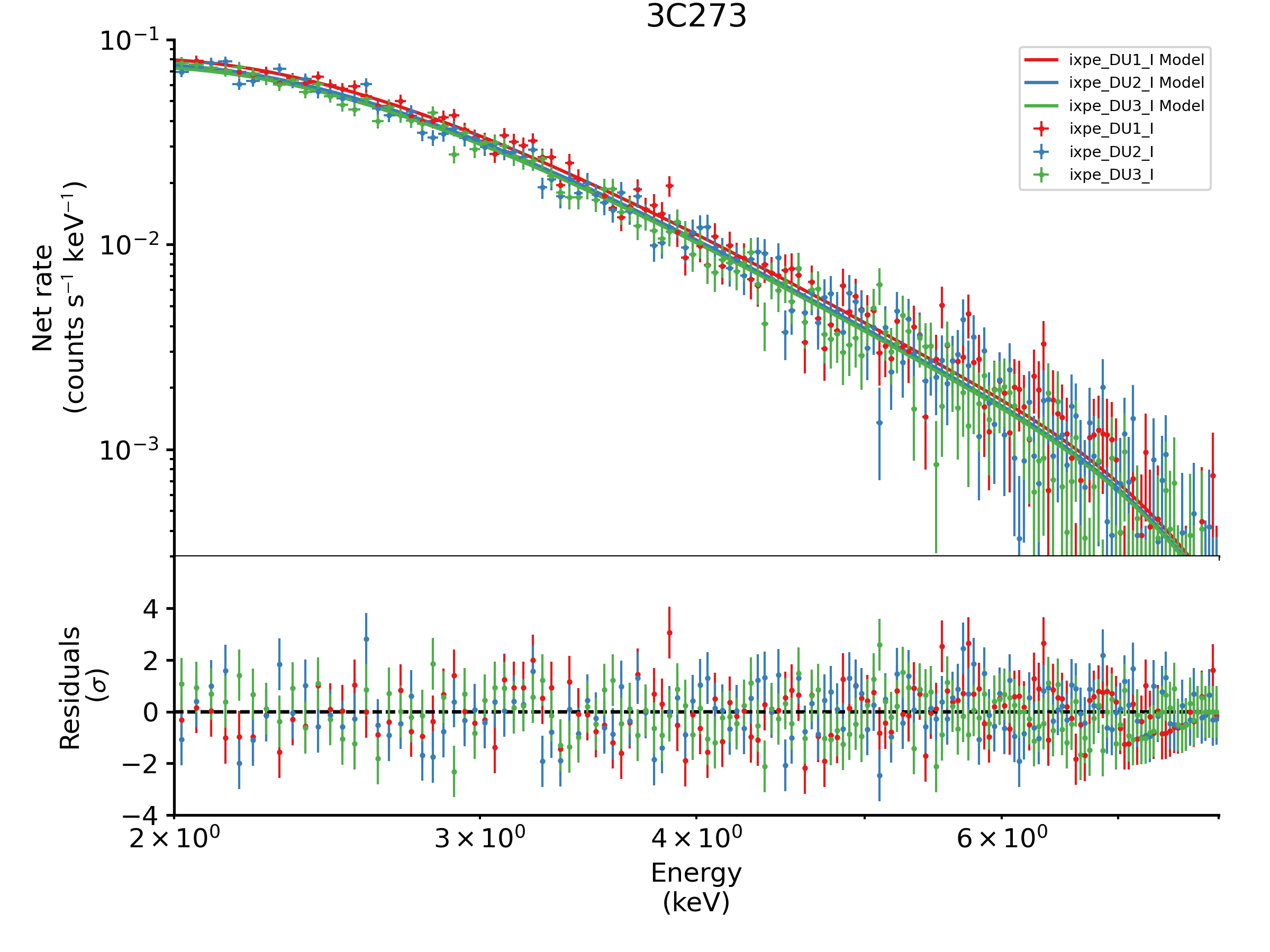}
 \includegraphics[width=8.cm]{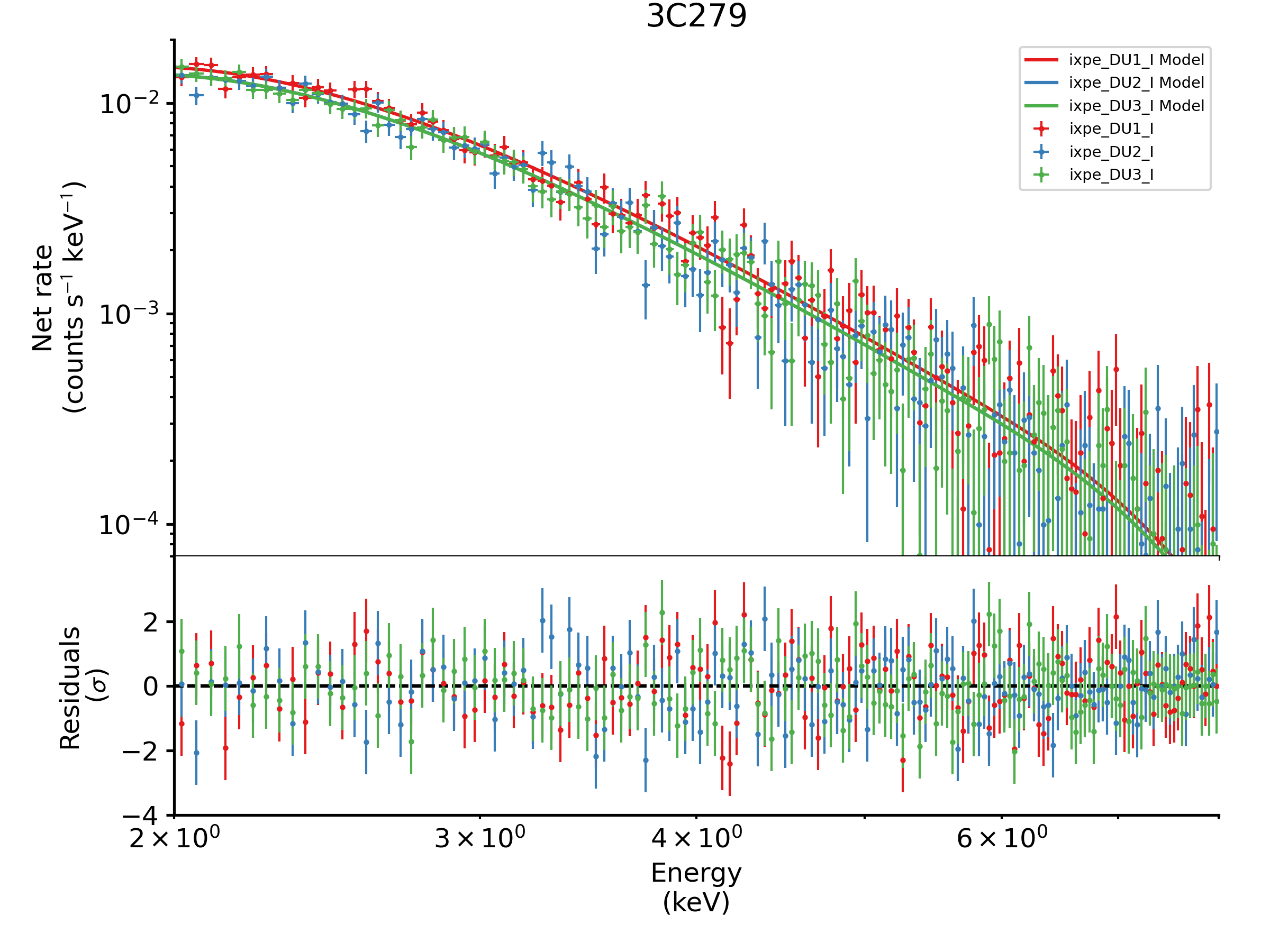}\\
 \includegraphics[width=8.cm]{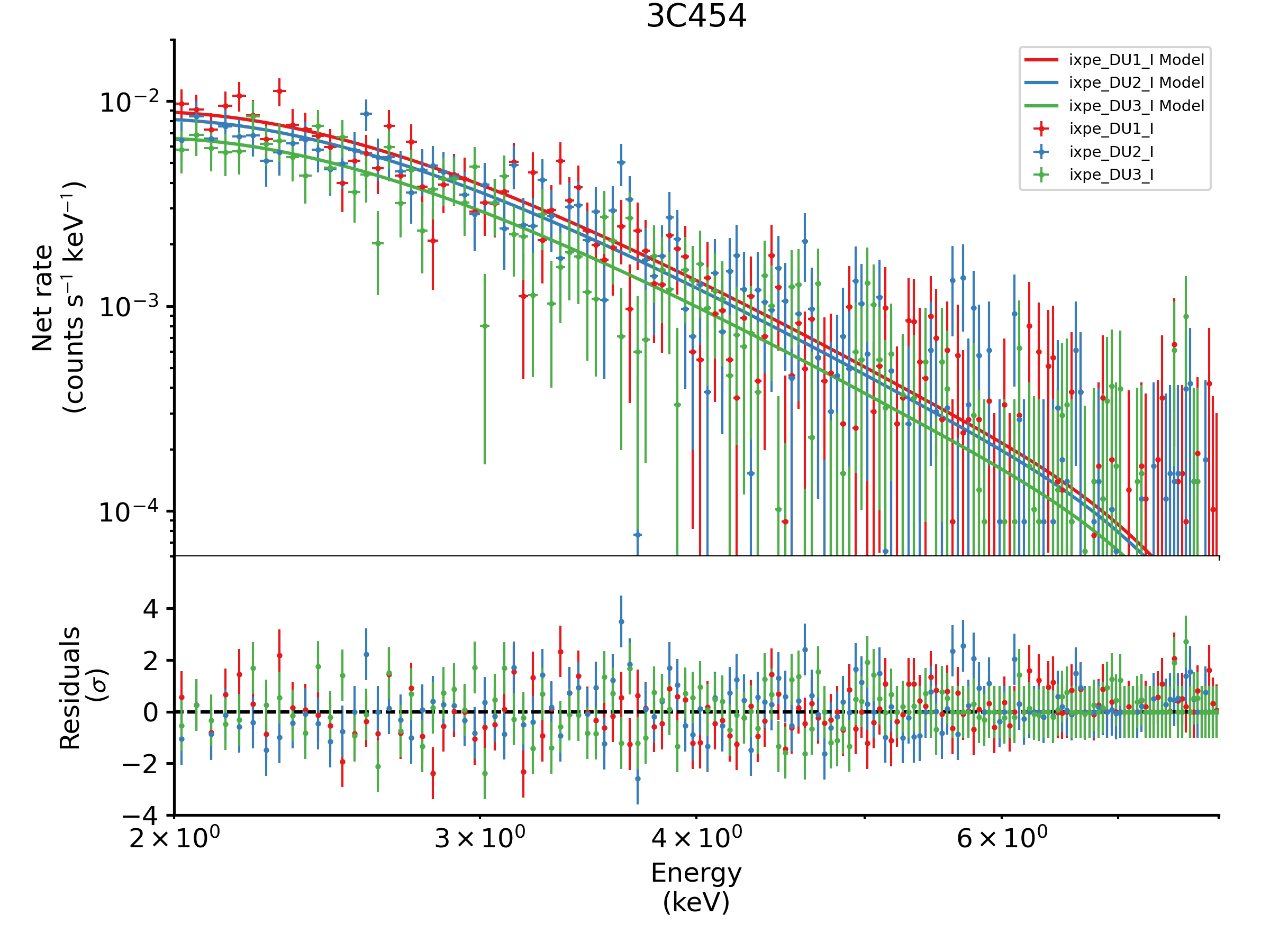}
 \includegraphics[width=8.cm]{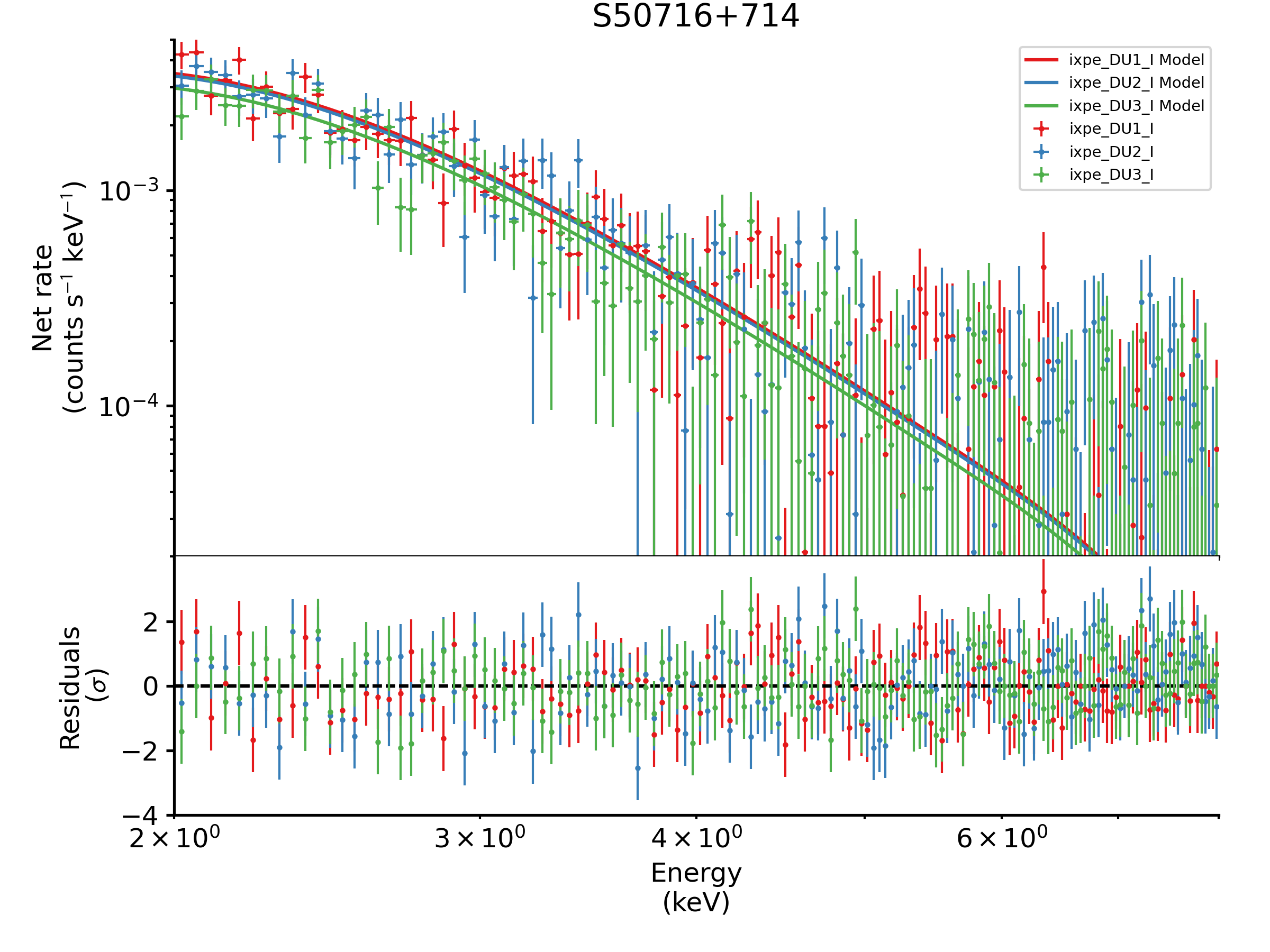}
 \caption{Background-subtracted Stokes $I$ spectra for the four sources
 (from top-left to bottom-right: 3C~273, 3C~279, 3C~454.3, and S5~0716+714). In each plot, the solid lines show the best fit model (absorbed power law), and the bottom panels illustrate the residuals of the data compared to the best fit model.}
\label{fig:Ispec}
\end{figure}

\begin{figure}[t]
\centering
 \includegraphics[width=8.cm]{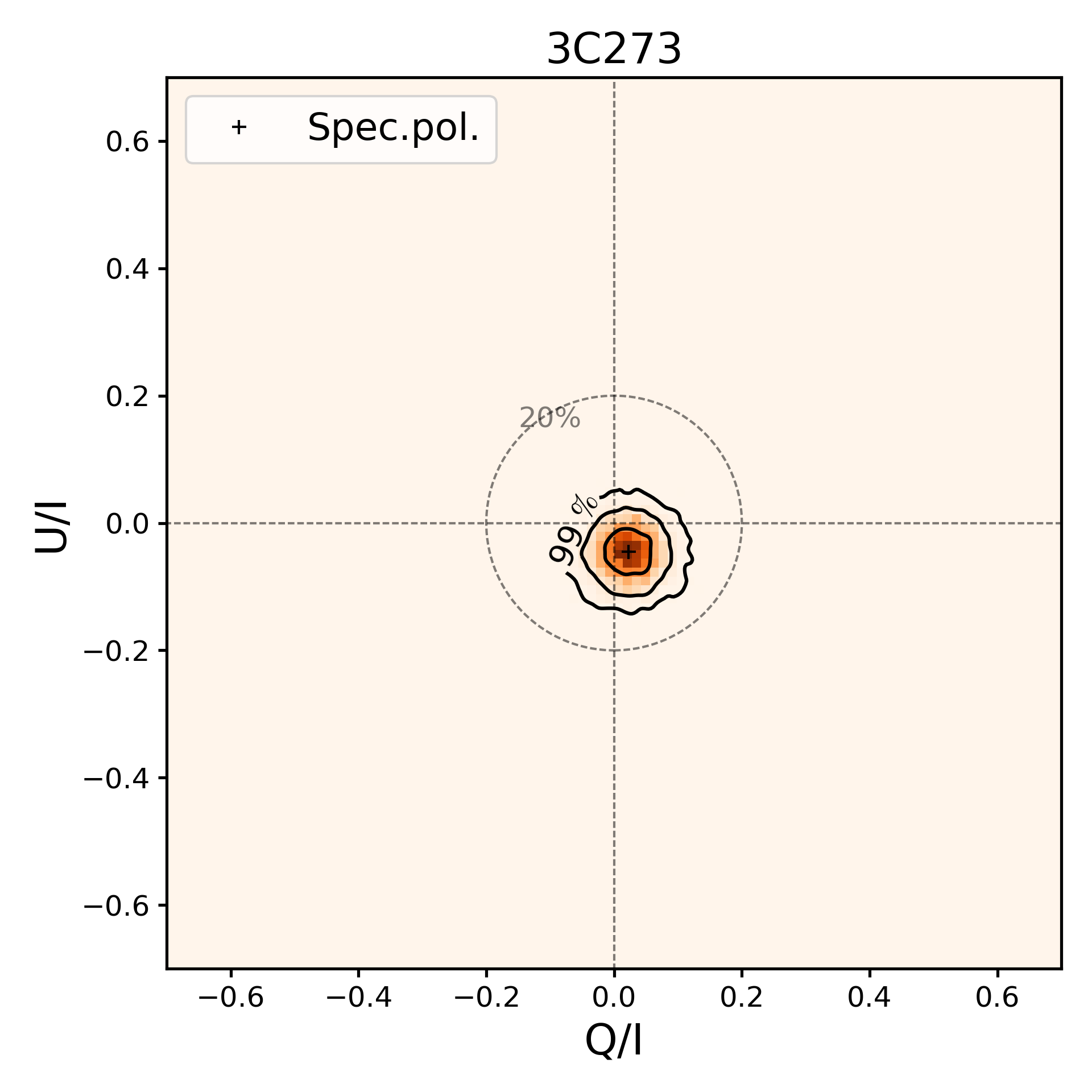}
 \includegraphics[width=8.cm]{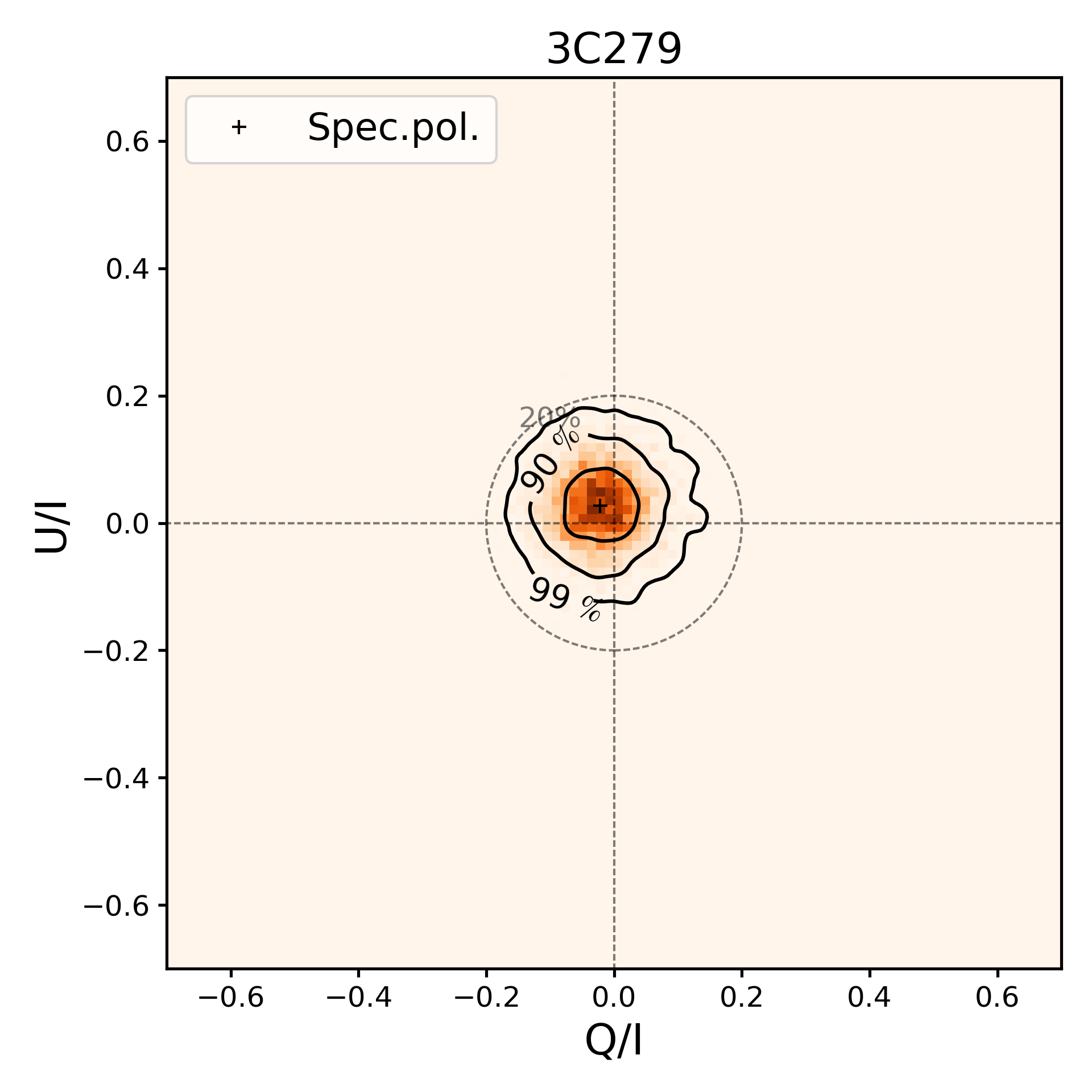}\\
 \includegraphics[width=8.cm]{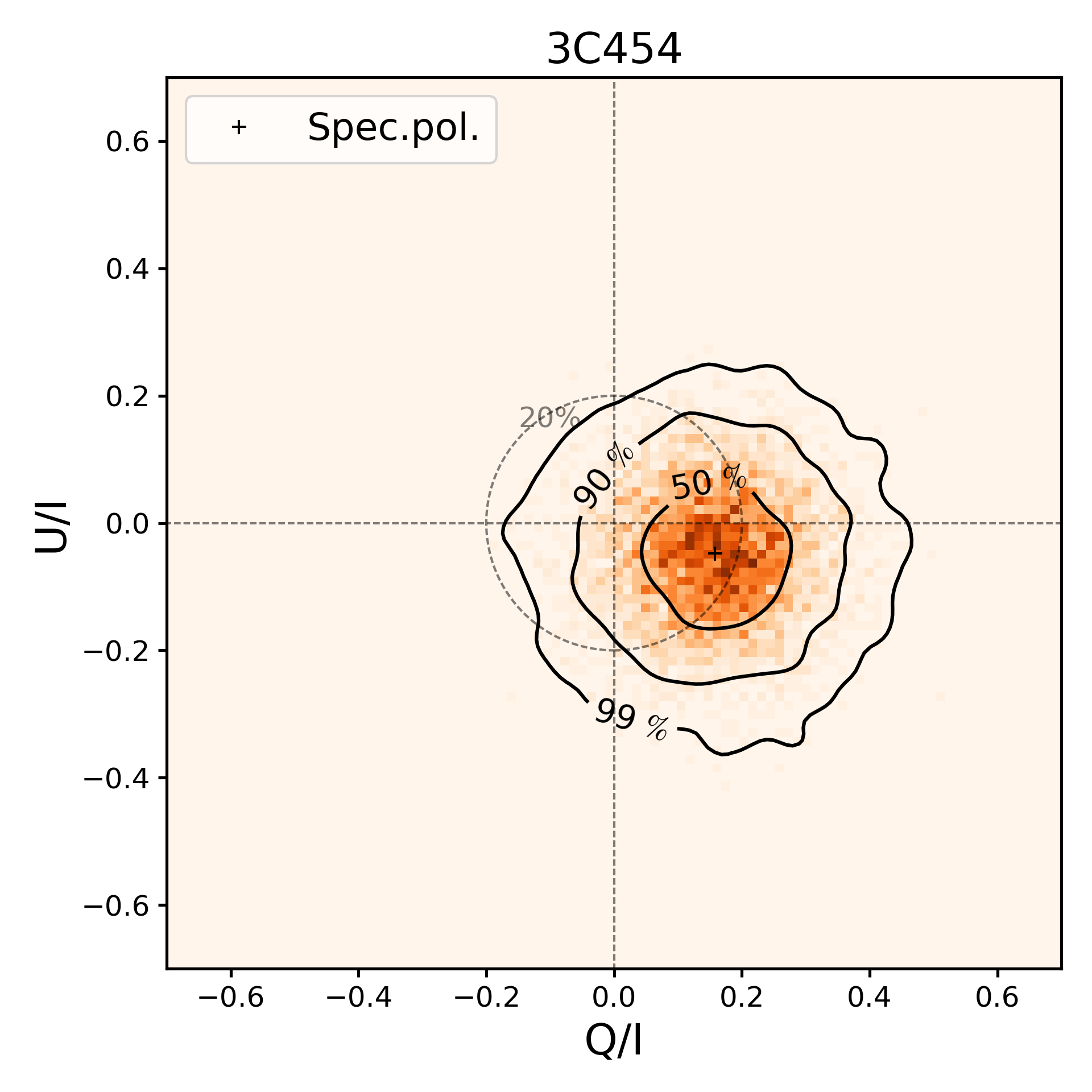}
 \includegraphics[width=8.cm]{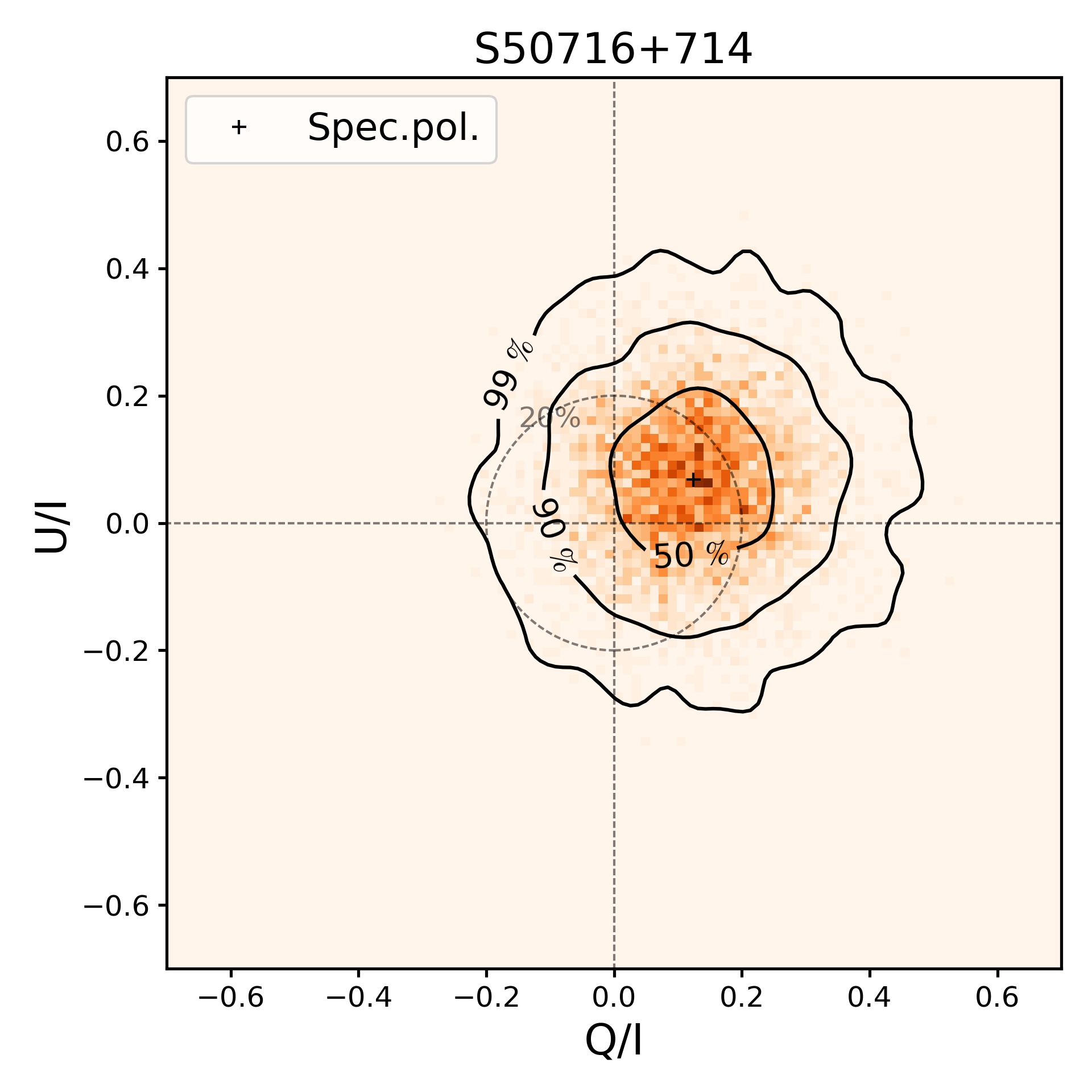}
 \caption{Stokes $Q/I$ and $U/I$ contour plots resulting from the spectropolarimetric fit for the four sources (from top-left to bottom-right: 3C~273, 3C~279, 3C~454.3, and S5~0716+714).  Confidence levels of 50\%, 90\%, and 99\% are shown and the dotted circles indicate the locus where the polarization fraction is 20\%.}
\label{fig:quContours}
\end{figure}

\begin{figure}[t]
\centering
 \includegraphics[width=8.cm]{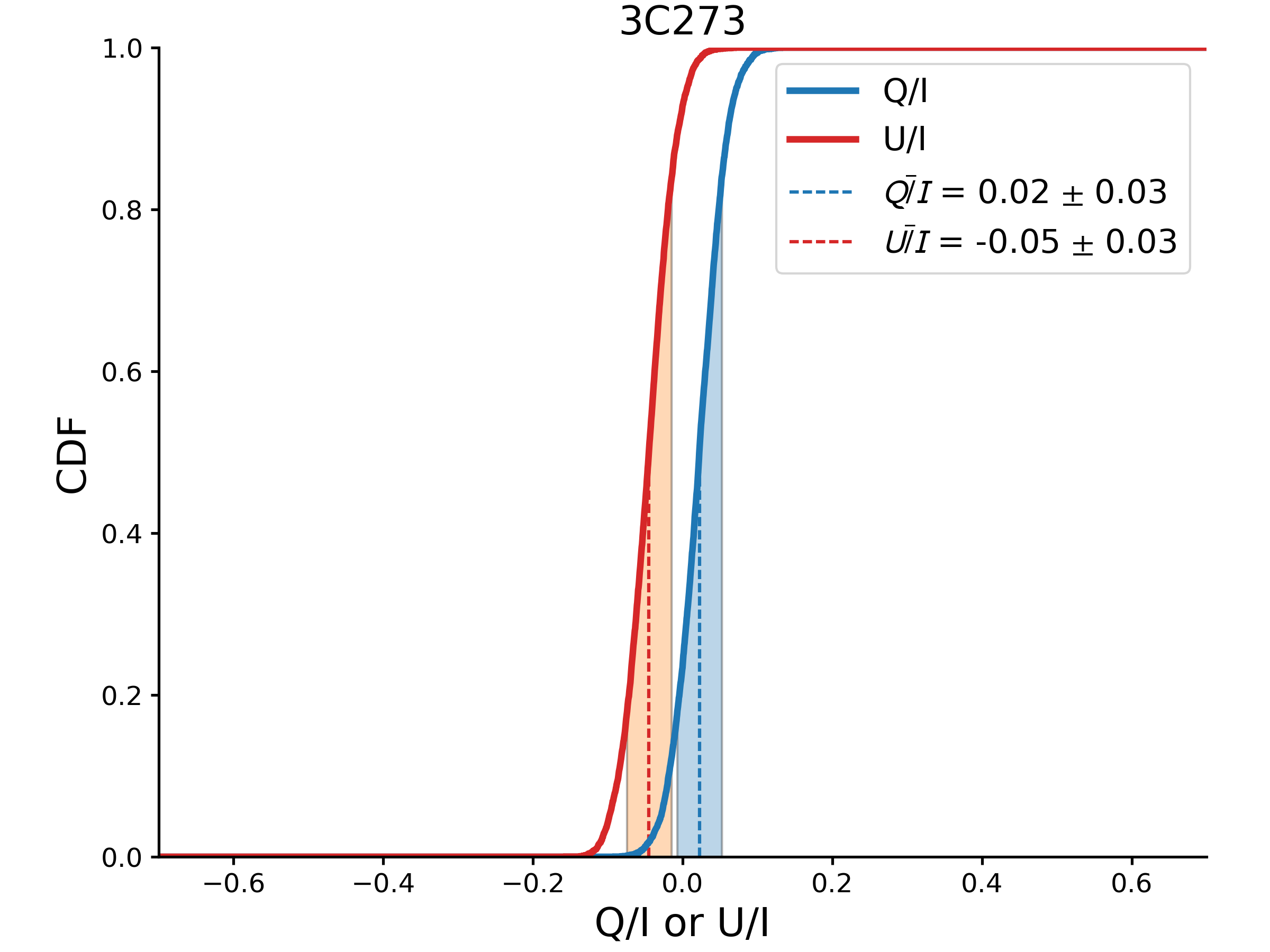}
 \includegraphics[width=8.cm]{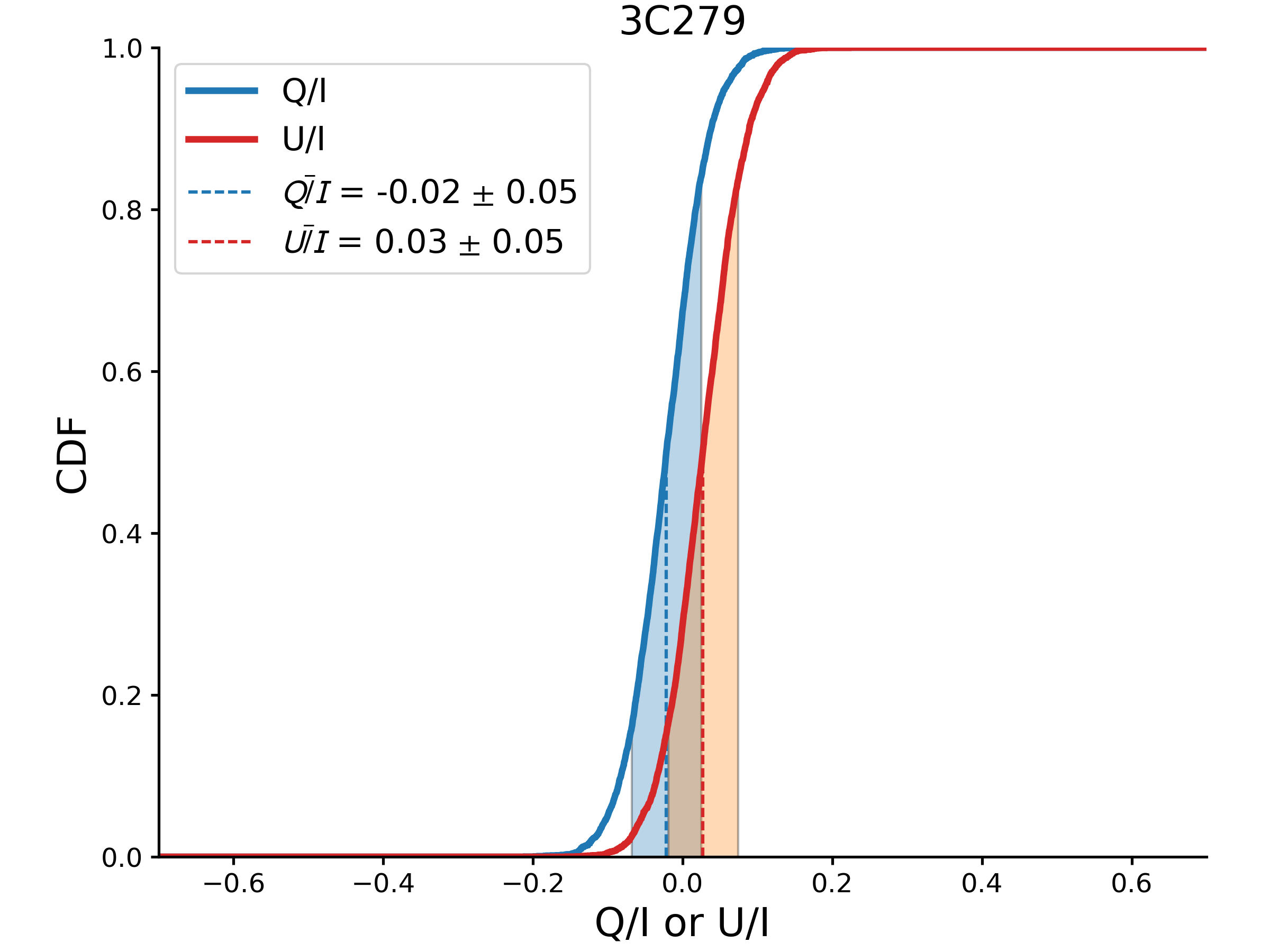}\\
 \includegraphics[width=8.cm]{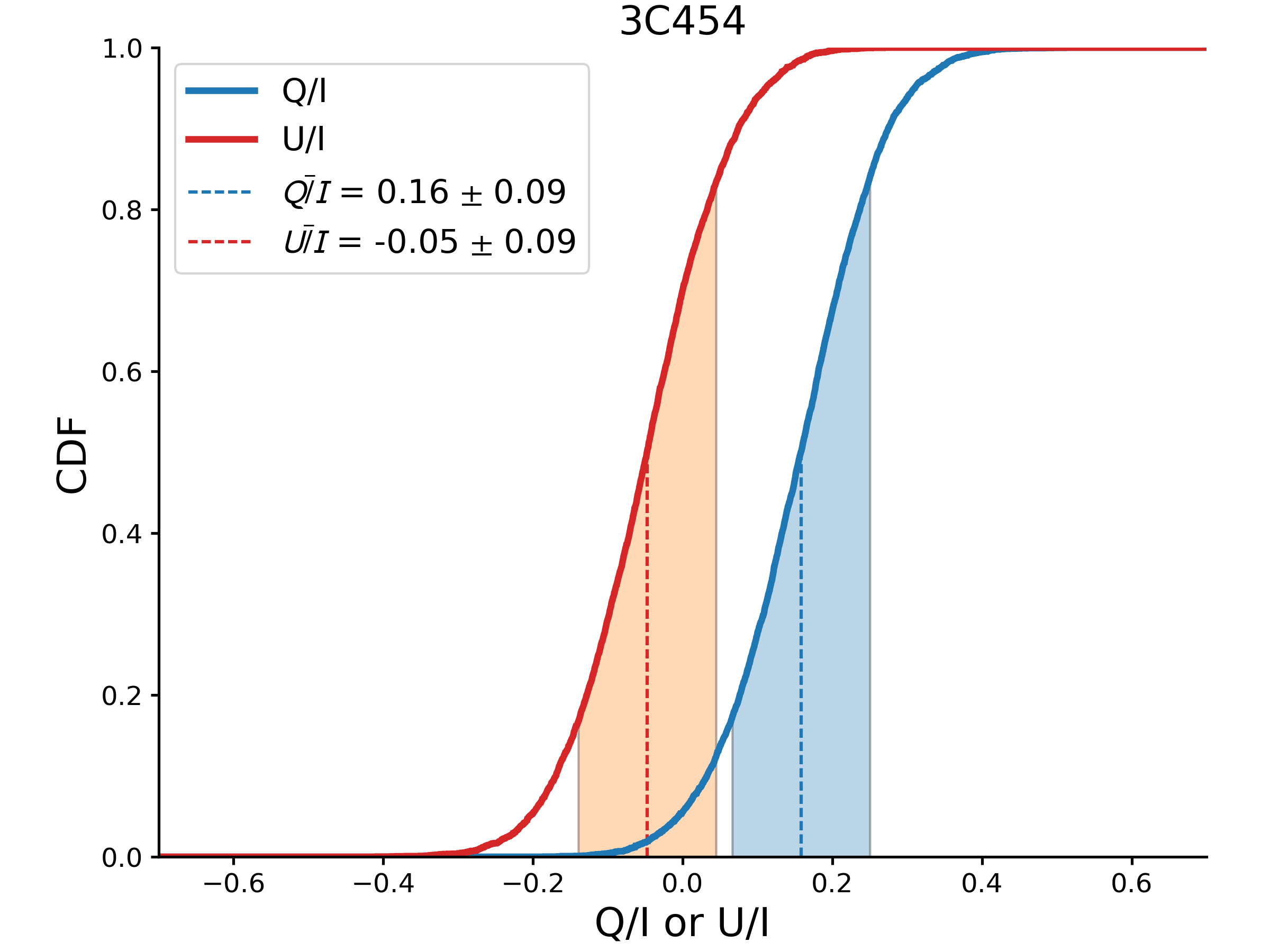}
 \includegraphics[width=8.cm]{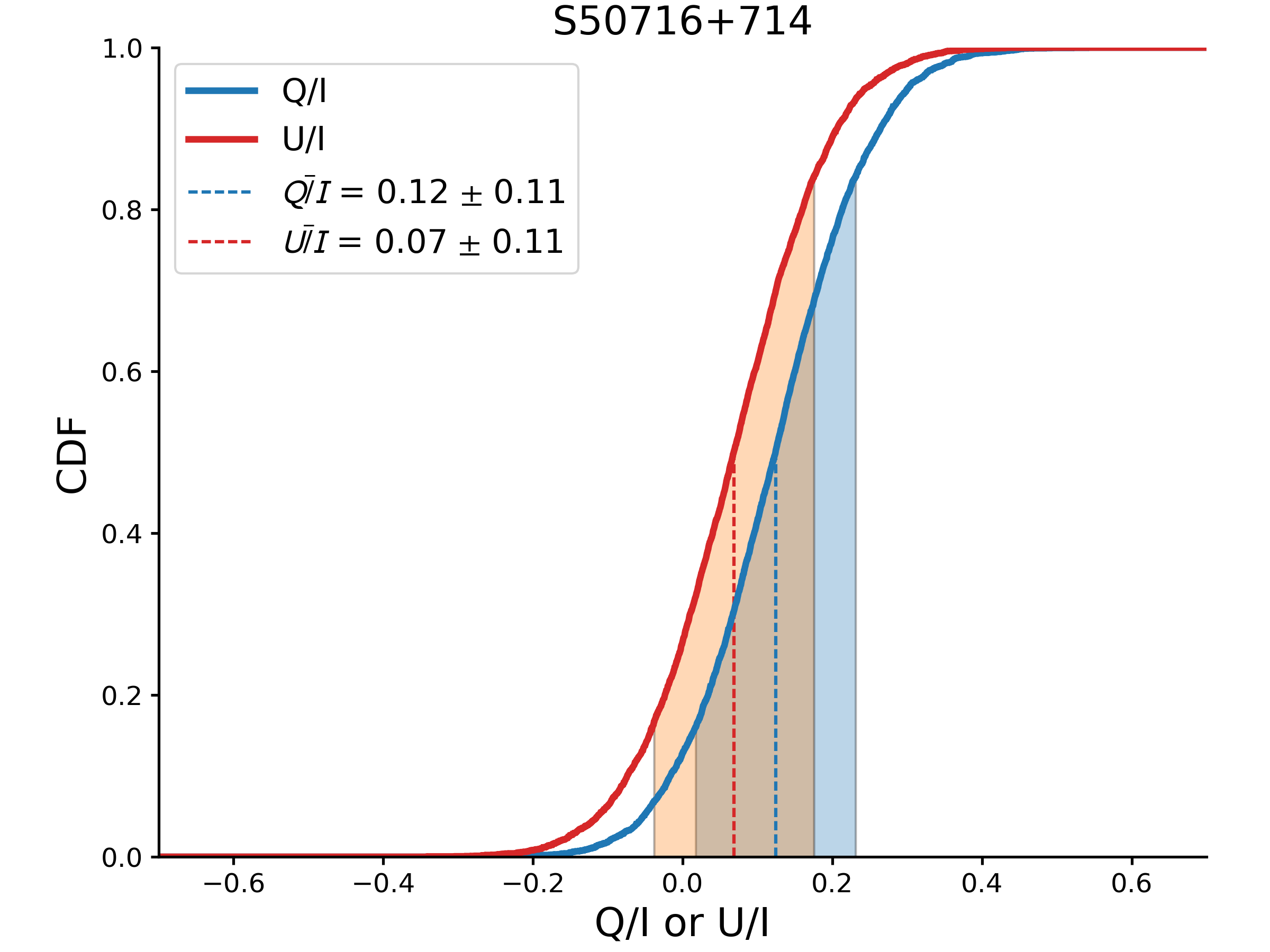}
 \caption{Stokes-Q and -U cumulative distribution functions (CDFs) for the best fit distributions for the four sources (from top-left to bottom-right: 3C~273, 3C~279, 3C~454.3, and S50716+714).}
\label{fig:quCDF}
\end{figure}

Spectropolarimetric fits were performed with the Multi-Mission Maximum Likelihood (\threeml) framework\footnote{\url{https://threeml.readthedocs.io/en/stable/index.html}} \citep{3ml}. In the 2--8 keV energy range, the spectra for the four sources are well described by an absorbed power law. The absorption column densities were fixed to the nominal Galactic values,\footnote{\url{https://heasarc.gsfc.nasa.gov/cgi-bin/Tools/w3nh/w3nh.pl}} and the polarization parameters are assumed to be constant through each observation and independent of energy.  Table~\ref{tab:3mlres} summarizes the best-fit parameters from the fits to each source and Fig.~\ref{fig:Ispec} shows the spectral fits.  Fig.~\ref{fig:quContours} and Fig.~\ref{fig:quCDF} show the polarization parameters derived from the spectropolarimetric fits.  An analysis using {\tt xspec} gave similar results.

\begin{deluxetable}{rcccc}[b]
\caption{Summary of Spectropolarimetric Results}
\label{tab:3mlres}
\tablehead{
\colhead{Source} & \colhead{N$_{\rm H}$} & \colhead{$\Gamma$} & \colhead{$K$} & \colhead{$\Pi_X$\tablenotemark{a}} \\[-0.2cm]
 &  \colhead{ ($10^{20}$ cm$^{-2}$)} & &
 \colhead{ (\rm $10^{-3}$ ph cm$^{-2}$ s$^{-2}$ keV$^{-1}$)} &  }
\startdata
3C 273      & 1.68 & $1.80\pm0.02$ & 12.1 & $<$12.8\% \\
3C 279      & 2.25 & $1.79\pm0.04$ & 2.28 & $<$15.5\% \\
3C 454.3    & 6.81 & $1.71\pm0.08$ & 1.22 & $<$40.3\% \\
S5 0716+714 & 2.88 & $2.29\pm0.20$ & 0.72 & $<$41.8\% \\
\enddata
\tablenotetext{a}{The 99\% confidence upper limits on $\Pi_X$ are computed using the $Q$ and $U$ best-fit Gaussian 1$\sigma$ errors and the best fit $Q/I$ and $U/I$ values.  See Sec.~\ref{sec:ixpe_spectr} for details.}
\end{deluxetable}

\section{Notes on Individual Sources}\label{sec:notes}

Here we give information regarding the individual sources and our contemporaneous multiwavelength campaign during the IXPE observations. The results from the multiwavelength campaigns are summarized in Appendix \ref{sec:multi} and in
Figs. \ref{fig:OIR_3C273}, \ref{fig:OIR_3C279}, \ref{fig:OIR_3C454}, and \ref{fig:OIR_0716}, and Tables \ref{tab:mult_obs_3C273}, \ref{tab:mult_obs_3C279}, \ref{tab:mult_obs_3C454}, and \ref{tab:mult_obs_0716}.

\subsection{3C~273}

3C273 ($z=0.158339$) is
one of the brightest X-ray emitting LSP blazars.  It is
a flat spectrum radio loud quasar with a strong ``Big Blue Bump'' that is interpreted as optical emission from an accretion disk \citep{Shields1978,Malkan1983}.
Consistent with the interpretation that the optical light is of thermal origin, it is notoriously unpolarized \citep{Stockman1984,Fernandes2020,Blinov2021}; hence it is not surprising that the polarization in the BVRI bands was $<0.5\%$ during our contemporaneous optical/infrared campaign. Complicating the interpretation of our results is that the X-ray emission in 3C~273 is most likely a mix of hot corona and jet emission \citep{Grandi2004,Chidiac2016}. The non-detection ($<9\%$) of X-ray polarization prevents us from discerning between these two possible origins of the X-ray emission.

\subsection{3C~279}

3C~279 ($z=0.538$) is
one of the most rapidly varying blazars, showing minute-timescale variability in $\gamma$-rays \citep{Ackermann2016}. It is also known to show large optical EVPA variations \citep{Kiehlmann2016} that might also be connected to repeating patterns of $\gamma$-ray activity \citep{Blinov2021-II}. During the IXPE observation the source polarization was about 4\% in millimeter(mm)-radio and 12\% (within uncertainties) in optical/IR (BVRIH). The EVPA showed a mild decrease from $\sim 175^\circ$ to $\sim165^\circ$ in the optical-IR bands. In a hadronic scenario, X-ray polarization should be stable -- less variable than the optical polarization \citep{Zhang2016}. Given the low-amplitude, slow EVPA variability in optical, and the fact we do not find evidence for significant variations of the X-ray EVPA (see Fig. \ref{fig:rotsearch}), we can exclude temporal depolarization effects. Thus, the fact that the optical/IR polarization degree is comparable to the X-ray upper limit ($<12.7\%$) strongly disfavors models predicting higher X-ray polarization than the optical, such as for Mrk 421 \citep{DiGesu2022,DiGesu2023} and Mrk 501 \citep{2022Natur.611..677L}.

\subsection{3C~454.3}

3C~454.3 ($z=0.859$) 
has had several well-studied
outbursts observed across a wide swath of the electromagnetic spectrum \citep[e.g.,][]{Jorstad2010,Jorstad2013,Weaver2019,Liodakis2020} and showed peculiar jet behavior \citep[e.g.,][]{Traianou2022}.  It showed large variation in the value of $\rm \Pi_O$ that can reach as high as 30\% \citep{Liodakis2020}. During the IXPE observation the source was at best weakly polarized. In mm-radio, we obtained only upper limits of $<$0.8\% at 3mm and $<$3.3\% at 1.3mm. The optical (VRI) polarization is below 1\% with a stable EVPA within uncertainties. The unpolarized state in the optical in combination with the X-ray upper limit ($<28\%$) prevents us from coming to any conclusion regarding the emission processes in this source but it may be interesting to re-observe this target when $\rm \Pi_O$ is much higher.

\subsection{S5~0716+714}

S5 0716$+$714 ($z=0.3$)
is a highly variable source that often shows intra-night variability \citep[e.g.,][]{Gupta2012}, large outbursts across the electromagnetic spectrum and TeV emission \citep[e.g.,][]{MAGIC2018}. During the IXPE observation, the source was observed in several optical and infrared bands (BVRIJHK) showing highly variable $\rm \Pi_O$ between 1-13\%. There is a large change in the optical EVPA before the IXPE observations (from $\sim125^\circ$ to $\sim50^\circ$). During the observation we observe a slow optical EVPA increase from $\sim50^\circ$ to $\sim100^\circ$. Unfortunately the source was in a quiescent X-ray flux state, providing an upper limit to the X-ray polarization of $<$26\%. Given the level of optical/IR polarization, we can only exclude models that predict several factors higher X-ray polarization than optical.  We note that the X-ray spectral index is steeper for this source than the others presented here, perhaps an indication of its ISP nature that might provide a more significant X-ray polarization detection during a soft X-ray flare.

\section{Summary}\label{sec:disc}

We presented results from the first year observations of LSP and ISP sources by IXPE, namely  3C~273, 3C~279, 3C~454.3, and S5~0716$+$714. All the IXPE observations were supplemented with a contemporaneous radio/optical/IR polarization campaign.  None of the sources yielded a significant ($>3\sigma$) X-ray polarization detection.  Instead, we are able to constrain $\rm \Pi_X$ in the 2--8~keV band to be $< 13$\% for 3C~273 and 3C~279 and $< 28$\% for 3C~454.3 and S5~0716$+$714. The non-detection of X-ray polarization in these sources is consistent with previous IXPE results on BL Lac in both LSP and ISP states \citep{Middei2023,Peirson2023} as well as other radio galaxies \citep{Ehlert2022}.

For 3C~273 and 3C~454.3, the undetected X-ray polarization could point towards inverse-Compton scattering, either external Compton or SSC, however the unpolarized optical/IR emission during the IXPE observations is preventing us from coming to definitive conclusions. For S5~0716$+$714, the highly variable optical polarization degree and the high upper limit ($<26\%$) on the X-ray polarization makes any interpretation difficult. Our results disfavor a scenario where the X-ray polarization is a factor of several higher than the optical. Such a scenario can involve, for example, a pure proton synchrotron model \citep{Zhang2013,Paliya2018,Zhang2019} or scattering from relativistically moving plasma containing relatively cold electrons \citep{Begelman1987}.

On the other hand, 3C~279 shows fairly stable $\Pi$ and EVPA in the optical/IR that remains persistently high ($>10\%$) and the upper limit to the $\Pi_X$ that is comparable ($<12.7\%$). IXPE observations of BL Lac showed a similar picture, i.e., undetected X-ray polarization ($<16\%$) with $\rm \Pi_O$ that is comparable or exceeds the X-ray limits \citep{Middei2023,Peirson2023}. However, in the case of BL Lac the highly variable EVPA we observe in the optical could lead to depolarization of the otherwise highly polarized proton synchrotron emission making it consistent with the $\rm \Pi_X$ limits. In the case of 3C~279 we can exclude any such depolarization effects, therefore our results strongly disfavor a significant contribution from proton synchrotron to the emission.
While our results cannot yet differentiate between scattering or other hadronic processes, the emerging pattern of IXPE LSP/ISP observations suggests that $\Pi_X \lae 10\%$ and less than or comparable to the contemporaneous $\rm \Pi_O$.
Further IXPE observations can help further elucidate the X-ray emission mechanism in LSP/ISP blazars.

\begin{acknowledgments} 

The Imaging X-ray Polarimetry Explorer (IXPE) is a joint US and Italian mission.  The US contribution is supported by the National Aeronautics and Space Administration (NASA) and led and managed by its Marshall Space Flight Center (MSFC), with industry partner Ball Aerospace (contract NNM15AA18C).  The Italian contribution is supported by the Italian Space Agency (Agenzia Spaziale Italiana, ASI) through contract ASI-OHBI-2017-12-I.0, agreements ASI-INAF-2017-12-H0 and ASI-INFN-2017.13-H0, and its Space Science Data Center (SSDC) with agreements ASI-INAF-2022-14-HH.0 and ASI-INFN 2021-43-HH.0, and by the Istituto Nazionale di Astrofisica (INAF) and the Istituto Nazionale di Fisica Nucleare (INFN) in Italy.  This research used data products provided by the IXPE Team (MSFC, SSDC, INAF, and INFN) and distributed with additional software tools by the High-Energy Astrophysics Science Archive Research Center (HEASARC), at NASA Goddard Space Flight Center (GSFC).  Funding for this work was provided in part by contract 80MSFC17C0012 from the MSFC to MIT in support of the \ixpe project.  Support for this work was provided in part by the National Aeronautics and Space Administration (NASA) through the Smithsonian Astrophysical Observatory (SAO)
contract SV3-73016 to MIT for support of the Chandra X-Ray Center (CXC),
which is operated by SAO for and on behalf of NASA under contract NAS8-03060.
This research has made use of data from the RoboPol programme, a collaboration between Caltech, the University of Crete, IA-FORTH, IUCAA, the MPIfR, and the Nicolaus Copernicus University, which was conducted at Skinakas Observatory in Crete, Greece. The IAA-CSIC co-authors acknowledge financial support from the Spanish "Ministerio de Ciencia e Innovacion" (MCINN) through the "Center of Excellence Severo Ochoa" award for the Instituto de Astrof\'{i}sica de Andaluc\'{i}a-CSIC (SEV-2017-0709). Acquisition and reduction of the POLAMI, TOP-MAPCAR, and OSN data was supported in part by MICINN through grants AYA2016-80889-P and PID2019-107847RB-C44. Some of the data are based on observations collected at the Observatorio de Sierra Nevada, owned and operated by the Instituto de Astrof\'{i}sica de Andaluc\'{i}a (IAA-CSIC). Further data are based on observations collected at the Centro Astron\'{o}mico Hispano-Alem\'{a}n(CAHA), operated jointly by Junta de Andaluc\'{i}a and Consejo Superior de Investigaciones Cient\'{i}ficas (IAA-CSIC). The POLAMI observations were carried out at the IRAM 30m Telescope. IRAM is supported by INSU/CNRS (France), MPG (Germany) and IGN (Spain). Some of the data reported here are based on observations obtained at the Hale Telescope, Palomar Observatory as part of a continuing collaboration between the California Institute of Technology, NASA/JPL, Yale University, and the National Astronomical Observatories of China.
%This research made use of Photutils, an Astropy package for detection and photometry of astronomical sources (Bradley et al., 2019).
GVP acknowledges support by NASA through the NASA Hubble Fellowship grant  \#HST-HF2-51444.001-A awarded by the Space Telescope Science Institute, which is operated by the Association of Universities for Research in Astronomy, Inc., under NASA contract NAS5-26555. The data in this study include observations made with the Nordic Optical Telescope, owned in collaboration by the University of Turku and Aarhus University, and operated jointly by Aarhus University, the University of Turku and the University of Oslo, representing Denmark, Finland and Norway, the University of Iceland and Stockholm University at the Observatorio del Roque de los Muchachos, La Palma, Spain, of the Instituto de Astrofisica de Canarias. The data presented here were obtained in part with ALFOSC, which is provided by the Instituto de Astrof\'{\i}sica de Andaluc\'{\i}a (IAA) under a joint agreement with the University of Copenhagen and NOT. E.\ L.\ was supported by Academy of Finland projects 317636 and 320045. D.B., S.K., R.S., N. M., acknowledge support from the European Research Council (ERC) under the European Unions Horizon 2020 research and innovation programme under grant agreement No.~771282. CC acknowledges support by the European Research Council (ERC) under the HORIZON ERC Grants 2021 programme under grant agreement No. 101040021. The Dipol-2 polarimeter was built in cooperation by the University of Turku, Finland, and the Leibniz Institut f\"{u}r Sonnenphysik, Germany, with support from the Leibniz Association grant SAW-2011-KIS-7. This work was supported by JST, the establishment of university fellowships towards the creation of science technology innovation, Grant Number JPMJFS2129. This work was supported by Japan Society for the Promotion of Science (JSPS) KAKENHI Grant Numbers JP21H01137. This work was also partially supported by Optical and Near-Infrared Astronomy Inter-University Cooperation Program from the Ministry of Education, Culture, Sports, Science and Technology (MEXT) of Japan. We are grateful to the observation and operating members of Kanata Telescope. The research at Boston University was supported in part by National Science Foundation grant AST-2108622, NASA Fermi Guest Investigator grant 80NSSC22K1571, and NASA Swift Guest Investigator grant 80NSSC22K0537. This study used observations conducted with the 1.8 m Perkins Telescope Observatory (PTO) in Arizona (USA), which is
owned and operated by Boston University.
\end{acknowledgments}

\facilities{AZT-8, Calar Alto, IRAM-30m, \ixpe, LX-200, Kanata, Nordic Optical Telescope, Palomar, Perkins, Skinakas Observatory, Swift, T60, T90, T150}

\software{ {ixpeobssim \citep{2022SoftX..1901194B}, Astropy \citep{astropy:2013, astropy:2018},
Photutils \citep{Bradley2019}, 
NumPy and SciPy \citep{virtanen20}, IPython \citep{perez07}, HEASoft 6.30.1\footnote{http://heasarc.gsfc.nasa.gov/ftools} \citep{heasoft}}, \threeml\ \citep{3ml}}

\bibliographystyle{aasjournal} 
\bibliography{apj-jour,polarimetry22,software}

\appendix

\section{Multiwavelength polarization observations}\label{sec:multi}

\begin{figure*}
    \centering
     \includegraphics[width=\textwidth]{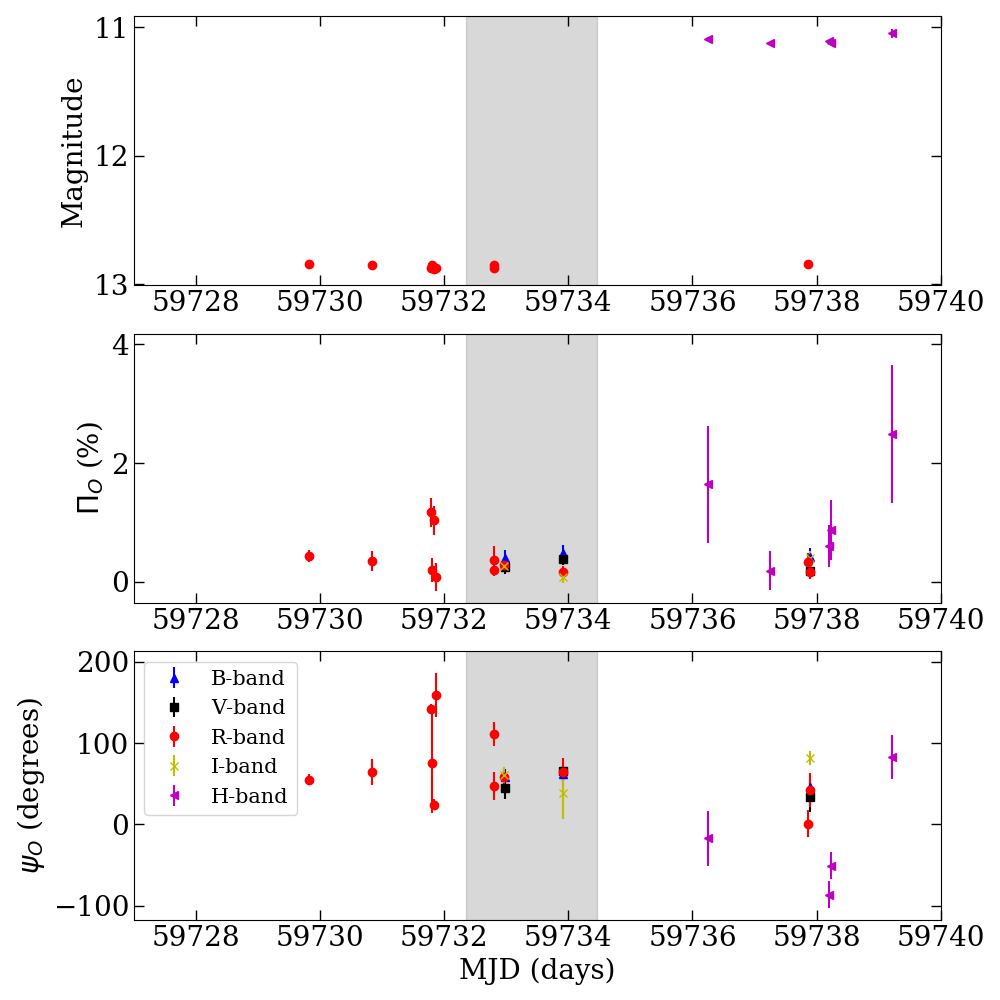}
    \caption{Contemporaneous optical and infrared observations of 3C~273. The top panel shows the brightness in magnitudes, middle panel the polarization degree, and bottom panel the EVPA. The grey shaded area marks the duration of the \ixpe observation.}
    \label{fig:OIR_3C273}
\end{figure*}

\begin{figure*}
    \centering
     \includegraphics[width=\textwidth]{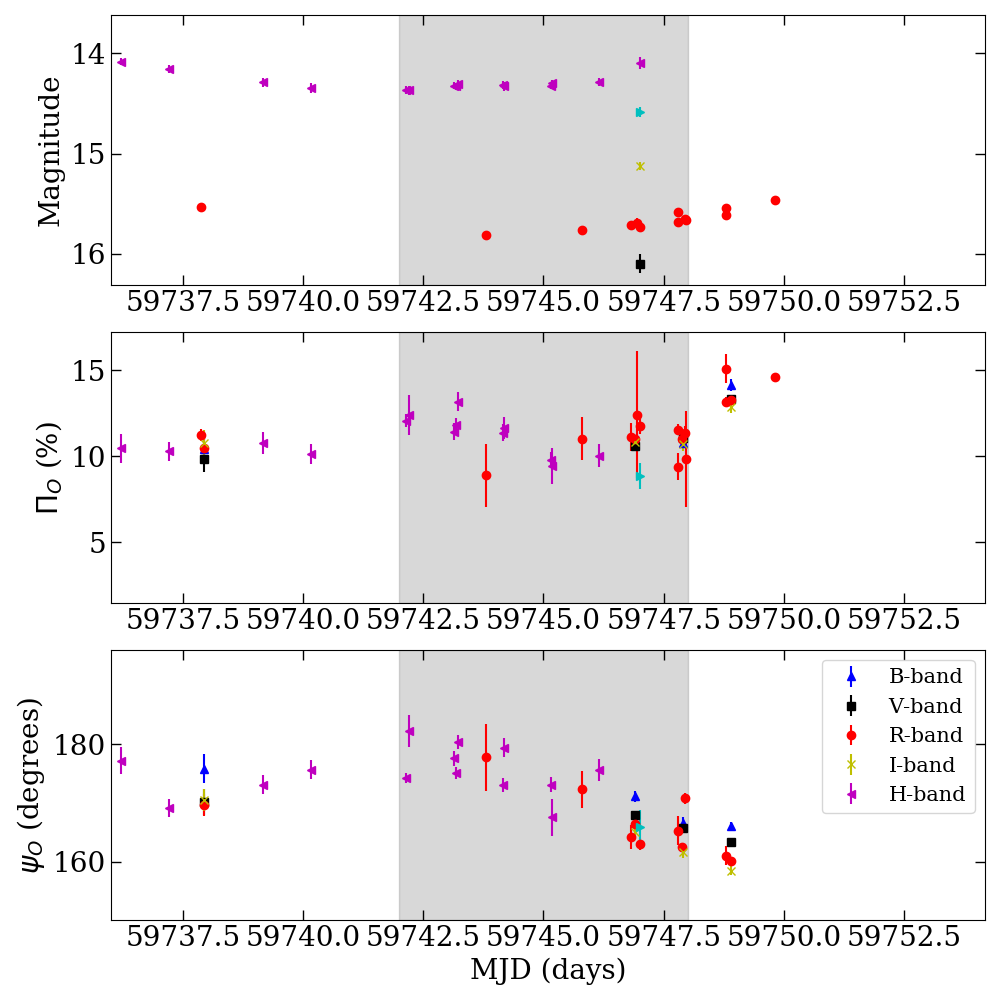}
    \caption{Same as Fig. \ref{fig:OIR_3C273} but for 3C~279.}
    \label{fig:OIR_3C279}
\end{figure*}

\begin{figure*}
    \centering
     \includegraphics[width=\textwidth]{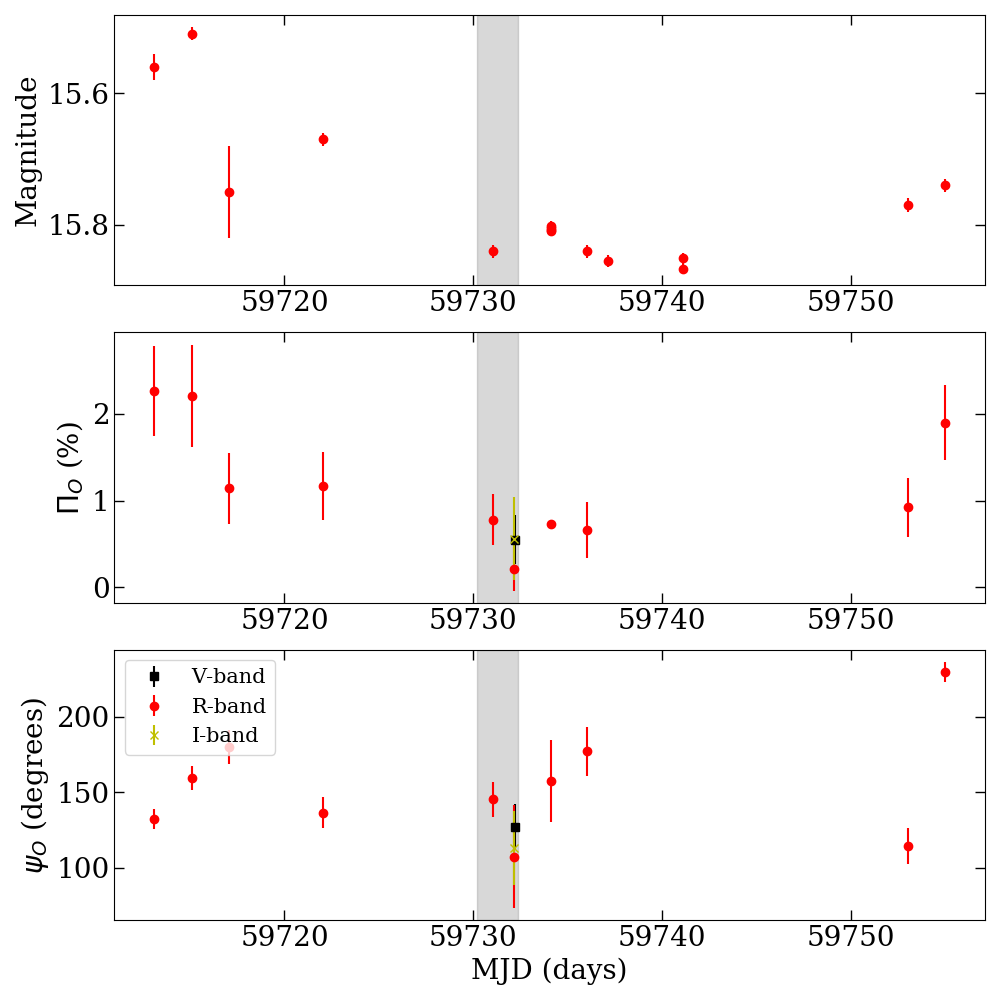}
    \caption{Same as Fig. \ref{fig:OIR_3C273} but for 3C~454.3.}
    \label{fig:OIR_3C454}
\end{figure*}

\begin{figure*}
    \centering
     \includegraphics[width=\textwidth]{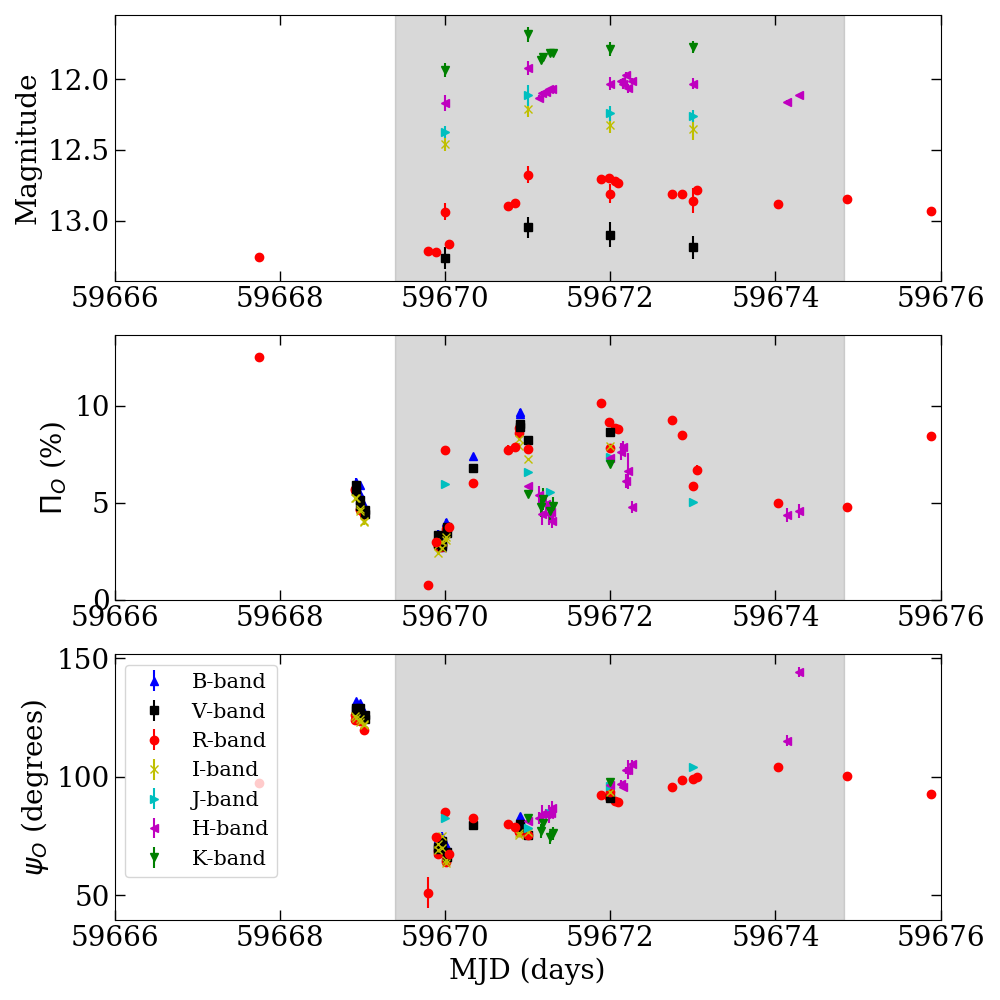}
    \caption{Same as Fig. \ref{fig:OIR_3C273} but for S5~0716+714.}
    \label{fig:OIR_0716}
\end{figure*}

Contemporaneous to the IXPE observations several telescopes across the world provided multiwavelength polarization observations from radio to optical. In the mm--radio spectral bands, all the sources were observed with the IRAM-30m telescope as part of the Polarimetric Monitoring of AGN at Millimeter Wavelengths (POLAMI\footnote{\url{http://polami.iaa.es/}}) project at 3.5 mm (86.24 GHz) and 1.3 mm (228.93 GHz) \citep{Agudo2018, Agudo2018-II, Thum2018}. S5~0716+714, 3C~273, and 3C~279 were observed in the J, H, and K infrared bands with using the 200-inch Palomar Hale telescope and the WIRC+Pol\footnote{\url{https://github.com/WIRC-Pol/wirc_drp}} instrument \citep{Tinyanont2019a,Tinyanont2019b,Millar-Blanchaer2021,Masiero2022}; the Kanata telescope using the Hiroshima Optical and Near-InfraRed camera (HONIR, \citealp{Kawabata1999,Akitaya2014}); and the IR camera MIMIR\footnote{https://people.bu.edu/clemens/mimir/index.html} \citep{Clemens2012} at the Perkins Telescope (PTO, Flagstaff, AZ). In the optical, all the sources we observed in B, V, R, I bands by the AZT-8 \& LX-200 telescopes (St. Petersburg University); Calar Alto and Sierra Nevada observatories; DIPOL-2 polarimeter at the Haleakala observatory T60 telescope \citep{Piirola1973,Kosenkov2017,Berdyugin2018,Berdyugin2019,2021AJ....161...20P}; HONIR at the Kanata telescope; Alhambra Faint Object Spectrograph and Camera (ALFOSC) at the Nordic Optical Telescope \citep{Hovatta2016,Nilsson2018}, and the RoboPol polarimeters at the Skinakas observatory \citep{Panopoulou2015,Ramaprakash2019,Blinov2021}. Figures \ref{fig:OIR_3C273}, \ref{fig:OIR_3C279}, \ref{fig:OIR_3C454}, and \ref{fig:OIR_0716} show the observations for the individual sources and Tables \ref{tab:mult_obs_3C273}, \ref{tab:mult_obs_3C279}, \ref{tab:mult_obs_3C454}, and \ref{tab:mult_obs_0716} the median values for the polarization degree and EVPA during the IXPE observations.

\begin{table}
\centering
\caption{Polarization Results Contemporaneous with the {\rm IXPE} observation of 3C~273. }
\begin{tabular}{lcccc}
\hline
Telescope & $\rm \Pi$  (\%)& $s_{\Pi}$ & $\rm \psi$ (deg.) & $s_{\psi}$ \\
\hline
POLAMI (3 mm) &  $5.55\pm1.17$  & 0.06 & $146\pm6$ & 0.3\\
POLAMI (1.3 mm) &  $5.23\pm0.8$ & 0.58 & $150\pm4$ & 6.6\\
AZT-8 \& LX-200 (R-band) & $0.38\pm0.23$ & -- & $111\pm14$ & -- \\
NOT (B-band)   & $0.45\pm0.13$ &0.05 & $60\pm9$ &1.45 \\
NOT (V-band)   & $0.31\pm0.11$ &0.07 & $550\pm11$ &10.5 \\
NOT (R-band)   & $0.23\pm0.1$ &0.05 & $61\pm14$ &3.4 \\
NOT (I-band)   & $0.18\pm0.1$ &0.09 & $50\pm21$ &11.1 \\
Skinakas (R-band)& $0.2\pm0.1$ & -- & $47\pm17$ & --\\
\hline
\end{tabular}
\tablecomments{The uncertainties for $\Pi$ and $\psi$ (i.e. EVPA) are either the uncertainty of the measurement or the median uncertainty in the case of multiple measurements. The quantities $s_{\Pi}$ and $s_{\psi}$ represent the standard deviations of $\Pi$ or $\psi$, respectively.}
\label{tab:mult_obs_3C273}
\end{table}

\begin{table}
\centering
\caption{Polarization Results Contemporaneous with the {\rm IXPE} observation of 3C~279. }
\begin{tabular}{lcccc}
\hline
Telescope & $\rm \Pi$  (\%)& $s_{\Pi}$ & $\rm \psi$ (deg.) & $s_{\psi}$ \\
\hline
POLAMI (3 mm) & $4.09\pm0.34$ & 0.15 & $359\pm2$ & 0.84\\
POLAMI (1.3 mm) & $<4.6$  & -- & -- & --\\
AZT-8 \& LX-200 (R-band)&  $10.2\pm1$ &0.97 & $169\pm3$ & 5.5\\
Calar Alto \& SNO (R-band)  & $11\pm2$ &3.4 & $154\pm6$ &10.6 \\
Kanata (R-band)& $11.75\pm0.42$ &-- & $163\pm1$ & --\\
Kanata (J-band)& $8.88\pm0.75$ & -- & $166\pm3$  & --\\
NOT (B-band) &  $10.77\pm0.35$ &0.03 & $169\pm1$ &2.26 \\
NOT (V-band) &  $10.86\pm0.26$ &0.23 & $167\pm0.7$ &1.1 \\
NOT (R-band) &  $11.01\pm0.26$ &0.01 & $164.4\pm0.6$ &2.04 \\
NOT (I-band) &  $10.79\pm0.36$ &0.07 & $163.3\pm1$ &1.72 \\
Perkins (H-band) & $11.54\pm0.5$ &1.13 & $175\pm1$ &4.06\\
Skinakas (R-band) & $11.55\pm0.33$ & -- & $160.4\pm0.8$ & --\\
\hline
\end{tabular}
\tablecomments{See the Table~\ref{tab:mult_obs_3C273} note.}
\label{tab:mult_obs_3C279}
\end{table}

\begin{table}
\centering
\caption{Polarization Results Contemporaneous with the {\rm IXPE} observation of 3C~454.3. }
\begin{tabular}{lcccc}
\hline
Telescope & $\rm \Pi$  (\%)& $s_{\Pi}$ & $\rm \psi$ (deg.) & $s_{\psi}$ \\
\hline
POLAMI (3 mm) &  $<0.82$  & -- & -- & --\\
POLAMI (1.3 mm) &  $<3.3$ & -- & -- & --\\
AZT-8 \& LX-200 (R-band) & $<1$ & -- &  -- & --\\
Calar Alto (R-band)  & $0.72\pm0.3$ &0.28 & $157\pm15$ &27 \\
NOT (B-band)  & $<1$ & -- & -- & -- \\
NOT (V-band)  & $0.55\pm0.3$ & -- & $127\pm15$ & -- \\
NOT (R-band)   & $0.21\pm0.25$ & -- & $107\pm34$ & -- \\
NOT (I-band)   & $0.56\pm0.48$ & -- & $113\pm25$ & -- \\
Skinakas (R-band) & $0.67\pm0.33$ & -- & $145\pm12$ & --\\
\hline
\end{tabular}
\tablecomments{See the Table~\ref{tab:mult_obs_3C273} note.}
\label{tab:mult_obs_3C454}
\end{table}

\begin{table}
\centering
\caption{Polarization Results Contemporaneous with the {\rm IXPE} observation of S5 0716+714. }
\begin{tabular}{lcccc}
\hline
Telescope & $\rm \Pi$  (\%)& $s_{\Pi}$ & $\rm \psi$ (deg.) & $s_{\psi}$ \\
\hline
POLAMI (3 mm) &  $<1.1$  & -- & -- & --\\
POLAMI (1.3 mm)  &  $<3.8$ & -- & -- & --\\
AZT-8 \& LX-200 (R-band) & $7.86\pm0.17$ &2.76 & $89.7\pm0.6$ & 14.24\\
Kanata (R-band)&  $7.75\pm0.04$ & 0.83 &$89.2\pm0.13$ & 9.01\\
Kanata (J-band)& $6.28\pm0.06$ & 0.84 &$89.4\pm0.2$  & 10.24\\
NOT (B-band) & $3.59\pm0.18$ &2.74 & $74\pm1.4$ &5.05 \\
NOT (V-band)   & $3.39\pm0.14$ &2.53 & $71.2\pm1.3$ &4.54 \\
NOT (R-band)   & $3.19\pm0.14$ &2.47 & $69\pm1.2$ &4.57 \\
NOT (I-band)   & $3.03\pm0.13$ &2.28 & $71\pm1.2$ &4.41 \\
Palomar (J-band) & $4.52\pm0.07$ & -- & $85\pm1$ & -- \\
Palomar (H-band) & $5.57\pm0.07$ &0 & $85\pm1$ & --\\
Perkins (H-band) & $4.86\pm0.38$ &1.26 & $96\pm2.2$ &16.95\\
Perkins (K-band) & $4.81\pm0.49$ &0.22 & $77\pm2.9$ &2.25\\
T60 (B-band) & $7.41\pm0.12$ & -- & $82.1\pm0.5$ &  -- \\
T60 (V-band) & $6.8\pm0.16$ & -- & $79.8\pm0.7$ &  -- \\
T60 (R-band) & $6.04\pm0.15$ & -- & $82.8\pm0.7$ &  -- \\
\hline
\end{tabular}
\tablecomments{See the Table~\ref{tab:mult_obs_3C273} note.}
\label{tab:mult_obs_0716}
\end{table}

\end{document}